\documentclass[11pt]{article}
\usepackage{amssymb,amsmath,amsfonts} 
\usepackage{comment}

\newtheorem{theorem}{Theorem}[section]

\newtheorem{lemma}[theorem]{Lemma}

\newcommand{\B}{\paramBB}
\newcommand{\lambdap}{{\Lambda_\omega} }
\newcommand{\discs}{{\mathcal D}}

\newcommand{\fdp}{{\langle\omega\rangle^\perp}}
\newcommand{\lomegar}{{\langle\omega\rangle}}
\def\norm#1{ |#1 |}
\def\Norm#1{ \|#1 \|}

\renewcommand{\Im}{\, {\rm Im}\,}

\newcommand\beq[1]{ \begin{equation}\label{#1} }
\newcommand{\eeq}{ \end{equation} }
\newcommand{\beqno}{ \[ }
\newcommand{\eeqno}{ \] }
\newcommand\beqa[1]{ \begin{eqnarray} \label{#1}}
\newcommand{\eeqa}{ \end{eqnarray} }
\newcommand{\beqano}{ \begin{eqnarray*} }
\newcommand{\eeqano}{ \end{eqnarray*} }

\newtheorem{definition}{Definition}[section]
\newcommand\dfn[1]{ \begin{definition}\label{#1} \rm}
\newcommand\edfn{ \end{definition} }
\newcommand{\proof}{\par\medskip\noindent{\bf Proof\ }}
\newcommand\equ[1]{{\rm (\ref{#1})}}
\newcommand{\nl}{{\smallskip\noindent}}
\newcommand{\giu}{{\medskip\noindent}}

\newcommand{\noi}{{\noindent}}
\newcommand\ugpereq[1]{ \stackrel{\equ{#1}}{=} }

\newcommand\lepereq[1]{ \stackrel{\equ{#1}}{\le} }

\newcommand{\dst}{\displaystyle}
\newcommand\Arccos{ {\, \rm Arccos\, }}
\newcommand{\real}{ {\mathbb R}   }
\def\Torus{{\mathbb T}}
\newcommand{\rn}{ {\real^n}   }
\newcommand{\p}{ {\pi}   }
\newcommand{\x }{ {\xi}   }

\renewcommand\subset{\subseteq}

\newcommand\pL{{\p_{{}_{L}}}}

\newcommand\pLd{{\p_{{}_{L_2}}}}

\newcommand\pLp{{\p_{{}_{L^\perp}}}}

\newcommand\bks{\backslash}

\newcommand\casesalt[3]{ \left\{  \begin{array}{ll}
 {#1} & \mbox{ {\rm if} ${#2}$} \\
 {#3} & \mbox{ {\rm otherwise\, .}}
 \end{array} \right.}

\newcommand\angolo{\,\angle\,}

\newcommand\idmap{{\rm Id}}

\newcommand\parama{\alpha}                        %  \delta              \alpha in Nekhoroshev
\newcommand\paramb{r}                             
\newcommand\paramc{\gamma}                 
\newcommand\paramd{R}                         
\newcommand\parame{\overline{\omega}}     
\newcommand\paramf{\widehat\omega}       
\newcommand\paramg{s}                               
\newcommand\paramh{\rho}                           
                         
\newcommand\paraml{\kappa}                                    
\newcommand\paramk{r}                                            
\newcommand\paramn{r_0}                             
\newcommand\paramo{R_0}                            
\newcommand\paramp{\delta}                           
\newcommand\paramq{q}                               
\newcommand\paramr{R}                                   
\newcommand\paramom{\underline{\omega}}  
\newcommand\params{t}                                   
\newcommand\paramB{U}                                
\newcommand\paramD{B}                                
\newcommand\paramBB{B}                              

\newcommand\parameps{\varepsilon}
\newcommand\paramepso{\varepsilon_*}
\newcommand\paramepsu{\varepsilon_0}

\title{
The Steep Nekhoroshev's Theorem
}

\begin{document}

\author{M. Guzzo 
\\ \footnotesize Dipartimento di Matematica Pura e Applicata 
\\ \footnotesize Universit\`a degli Studi di Padova 
\\ \footnotesize Via Trieste 63 - 35121 Padova, Italy
\and 
L. Chierchia 
\\ \footnotesize Dipartimento di Matematica e Fisica
\\ \footnotesize Universit\`a degli Studi Roma Tre
\\ \footnotesize Largo San L. Murialdo 1 - 00146 Roma, Italy
\\ \footnotesize 
\and 
G. Benettin
\\ \footnotesize Dipartimento di Matematica Pura e Applicata 
\\ \footnotesize Universit\`a degli Studi di Padova 
\\ \footnotesize Via Trieste 63 - 35121 Padova, Italy
}

\date{March 25,  2014}

\maketitle

\begin{abstract}
\noindent
Revising Nekhoroshev's geometry of resonances,
we provide a 
fully constructive and quantitative proof of
Nekhoroshev's  theorem for steep Hamiltonian systems proving, in particular, 
that the  exponential stability exponent 
can be taken to be  $1/ (2n \parama_1\cdots\parama_{n-2}$) ($\parama_i$'s being Nekhoroshev's steepness 
indices and $n\ge 3$ the number of degrees of freedom).
\end{abstract}

\section{Introduction and results}
\noindent
{\it {\bf A}. Motivations.} In 1977-1979 N.N. Nekhoroshev published a fundamental theorem (\cite{Nekhoroshev, Nekhoroshev79})
about the ``exponential stability'' (i.e., ``stability of action variables over times exponentially long with the inverse of the perturbation size'') of nearly--integrable, real--analytic Hamiltonian systems 
with Hamiltonian given, in standard action--angle coordinates, by   
\begin{equation}
H(I,\varphi)=h(I)+ \parameps f(I,\varphi )  ,\qquad (I,\varphi)\in \paramB\times {\Torus}^n\ ,
\label{Ham}
\end{equation}
where: $\paramB\subseteq {\mathbb R}^n$ is an open region, ${\mathbb T}^n={\mathbb R}^n/(2\pi {\mathbb Z})^n$ is the standard flat $n$--dimensional torus and 
 $\parameps$ is a 
small parameter. The integrable limit    $h(I)$ is assumed to satisfy a 
geometric condition, called  by Nekhoroshev 
``steepness''   (the definition is recalled in \equ{stdf} below). 
Under such assumptions, Nekhoroshev's states his theorem as follows\footnote{Compare \cite[p. 4 and p. 8]{Nekhoroshev}; see also\cite[p. 30]{Nekhoroshev} for a more detailed and precise statement.}:

\vskip 0.4 truecm\noindent
{\it Let $H$ in \equ{Ham} be real--analytic with $h$ steep. Then, there exist positive constants $a$, $b$ and 
$\paramepsu$
such that for any $0\leq \parameps<\paramepsu$ 
the solution $(I_t,\varphi_t)$
of  the (standard) Hamilton equations 
for
 $H(I,\varphi)$ satisfies 
 \begin{equation}
|I_t-I_0|\leq \parameps^{b}
\nonumber
\end{equation}
for any time $t$ satisfying
\begin{equation}
\nonumber
|t|\ \leq \  \frac{1}{\parameps}\ 
\exp\Big({\frac{1}{\parameps^{a}}}\Big)\  .
\end{equation}
Furthermore, $a$ and $b$ can be taken as follows:
\begin{equation}
a = \frac{2}{12\zeta + 3n+14}\ \ ,\ \quad
b = \frac{3 a}{2 \parama_{n-1}} 
\label{expa}
\end{equation}
where 
\begin{equation}
\zeta = \Big[ \parama_1\cdot \Big( \parama_2 \Big(\ldots \big(\parama_{n-3}(n \parama_{n-2}+
n-2)+n-3\big)+\ldots \Big)+2\Big) +1 \Big] -1  ,
\nonumber%
\label{expb}
\end{equation}
and $\parama_i$ are the steepness indices of $h$.
}

\vskip 0.4 truecm\noindent      
Usually, $a$ and $b$ are called the ``stability exponents''. Clearly, the most relevant 
quantity in this theorem is the  stability exponent $a$ appearing in the exponential, which 
gives the dominant time--scale for the stability of the action variables. 
The exponential stability exponent $a$
depends only on the number $n$ of degrees of freedom and on the values of 
the first $n-2$ steepness indices $\parama_i$, $i\leq n-2$. 
Notice that, for any fixed $n$, the ``best" exponents $a,b$ in \equ{expa}
are obtained in the 
special case $\parama_1=\ldots =\parama_{n-1}=1$, corresponding to 
convex (or quasi--convex) $h(I)$ (which is the simplest instance of steep function).
Actually, for any values of the steepness indices $\parama_i$, 
the parameter $\zeta$ defined in (\ref{expb}) grows faster than\footnote{For any fixed sequence $\parama_{j}\geq 1$, 
$j=1, 2, \ldots$, by considering a sequence of steep Hamiltonians $h_n$ with 
$n$ degrees of freedom and steepness indices $\parama_1,\ldots ,\parama_{n-1}$, the 
sequence of corresponding parameters $\zeta_n:=\zeta$ satisfies 
$\zeta_n-\zeta_{n-1}\geq (n-1) \parama_1\cdots \parama_{n-2}$, and the sequence 
of stability exponents $a_n:=a$ satisfies
$\displaystyle
a_{n}^{-1}-a_{n-1}^{-1}\geq 6 (n-1) \parama_1\cdots \parama_{n-2}$.} 
$n(n-1)/2 $.     
%\giu 
The hypotheses of Nekhoroshev's theorem, as pointed out by Nekhoroshev himself, are qualitatively optimal, and, in particular, non--steep Hamiltonian are in general non exponentially--stable \cite[\S 11]{Nekhoroshev}. Furthermore, Nekhoroshev proved that steepness is a generic (in $C^\infty$ category) property \cite{Nekhoroshev73}. Finally, several interesting problems (e.g., in Celestial Mechanics, compare below) are steep but do not satisfy simpler 
assumptions (such as quasi--convexity). For all these reasons it seems natural and 
important to try to optimize the exponential  stability exponents, especially 
with respect to the number $n$ of the degrees of freedom which, in applications,  
typically range from $n=3$ (restricted three-body problems) 
up to several tens (planetary problems); this has been done, up to now, under 
simplifying assumptions but not in the general steep case. This paper is devoted to the 
general case. 

\giu
Before stating our result, let us briefly review the main extensions, applications and improvements concerning Nekhoroshev's theorem.

\giu
Various extensions have been discussed, so as to cover the degeneracies of the Hamilton function which are usually 
met in some important mechanical systems (fast rotations of the 
Euler--Poinsot rigid body \cite{rigid,giro1,giro2}; the planetary $N$--body problem \cite{Nekhoroshev,niedpla,CP11}; restricted three 
body problems \cite{CF96}, elliptic equilibria \cite{FGB, Ni98,  GFB,poschel99}).
Furthermore, 
steepness 
could be used, in non-convex systems,  to 
study the long--term stability
in problems such as
the Lagrangian equilibrium 
points L4-L5 of the restricted three body problem \cite{BFG}, asteroids 
of the Main Belt \cite{MG,PavG08, LED08} and the Solar System \cite{SLG}. 

\noindent
As far as improvements of the theoretical stability bounds (i.e., improvements on the stability
exponent $a$), quite complete results have been achieved in the special case of
convex and quasi--convex functions $h$:
the proof of the theorem has been significantly simplified 
(see \cite{gallavotti86,BGG,BG86})
and the stability exponent improved up to $a=(2n)^{-1}$, 
(\cite{LOCNei}, \cite{LOC}, 
\cite{poschel};  see \cite{BM} for exponents which are intermediate
between $a=(2n)^{-1}$ and $a=(2(n-1))^{-1}$): such exponents (in the convex case) are nearly optimal, compare \cite{Zhang}. These  improvements have 
been obtained by exploiting specific geometric properties 
of the convex and quasi--convex cases, which allow  to use   
conservation of energy in order to obtain  
topological confinement of the actions (\cite{BG86}). In fact, in the convex case, 
the 
analysis of the 
geometry of resonances, that is, the geometry of the 
manifolds 
$$\{I\in \paramB: k\cdot \omega(I)=0\}\ ,\quad {\rm with}\quad \omega(I)=\nabla h(I)\ {\rm and} \ k\in {\mathbb Z}^n\ ,
$$
 is greatly 
simplified, since the frequency map $I \mapsto \omega(I)$ is a diffeomorphism;
on the other hand, in the general steep case, 
the Hamilton function cannot be used anymore 
in order to obtain topological confinement, and the geometry of resonances 
is significantly more complicate, due to possible folds and other 
degeneracies of the frequency map.
Furthermore,  while new different proofs of  Nekhoroshev's theorem have appeared (compare
\cite{Ni07}, which is based on the method of simultaneous Diophantine approximations introduced in \cite{LOC}),
{\sl no improvements on the original Nekhoroshev's stability exponents, in the general steep case,  are yet available}\footnote{In the paper \cite{Nie} there is a statement 
concerning improved values for the stability exponents, however, the proof appears to have a serious gap
and such values are not justified; see \cite{Nie.cor}
.}.

\vskip0.4truecm\noindent
In this paper, we 
revisit and extend Nekhoroshev's geometric analysis 
obtaining, in particular, for\footnote{The cases $n\le 2$ are, in general, totally stable and therefore are not included in our analysis.} $n\ge 3$, 
the {\sl new 
stability exponents} $a=1/(2np_1)$ and $b=a/\parama_{n-1}$ with $p_1$ being the product of the first $(n-2)$ steepness indices.\\
The new stability exponents represent an essential
improvement with respect to Eq.  (\ref{expa}); in particular, the dependence of $a^{-1}$ on the number 
of degrees of freedom improves from quadratic to linear. It is also remarkable that, 
for $\parama_1=\ldots =\parama_n=1$ (quasi--convex case), we obtain the ``optimal'' stability exponents 
proved in (\cite{LOCNei}, \cite{LOC}, \cite{poschel}), without using the local inversion 
of the frequency map, nor the Hamiltonian as a Lyapunov function. \\
A  precise and fully 
quantitative formulation is given in the following paragraph.

\vskip 0.4 cm
\noindent
{\it {\bf B}. Statement of the result.} 
A $C^1$ function  $h(I)$ is said to be steep  in $\paramB\subseteq {\mathbb R}^n$ with steepness indices 
 $\parama_1,\ldots ,\parama_{n-1}\geq 1$ and (strictly positive) steepness 
coefficients $C_1,\ldots ,C_{n-1}$ and $\paramb$,
if $\inf_{I\in U}\Norm{\omega(I)}>0$ and,  
for any $I\in \paramB$, for any  $j$--dimensional linear subspace  
$\Lambda \subset{\mathbb R}^n$  orthogonal to $\omega (I)$ 
 with $1\le j\leq n-1$, one has\footnote{For any vector $u\in {\mathbb C}^n$ we denote by $\Norm{u}:=\sqrt{\sum_i |u_i|^2}$ its hermitean 
norm and by $\norm{u}=\sum_i \norm{u_1}$.}
\begin{equation}
\max_{0 \leq \eta \leq \xi}\ \min_{u\in \Lambda:\Norm{u}=\eta}
\Norm{\pi_\Lambda \omega (I+u)} \geq C_j {\xi}^{\parama_j}
\ \ \ \ \forall\ \xi \in (0,\paramb]  ,
\label{stdf}
\end{equation}
where $\pi_\Lambda$ denotes the orthogonal projection over $\Lambda$.

\noindent
To deal properly with initial data near the boundary, we will use the following notation: for any $\eta>0$ and any $D\subseteq {\mathbb R}^n$, we let 
$D-\eta:=\{ I\in D :\, \overline{\B(I,\eta)}\subseteq D\}$, 
where 
$$\B (I,\eta)=\{ I'\in {\mathbb R}^n: \Norm{I'-I}<\eta\}$$ 
is the real euclidean  ball centered in $I$ of radius $\eta$ and $\overline{\B(I,\eta)}$ its closure.

\vskip 0.4 truecm 
\noindent
{\bf Theorem 1.} {\it 
Let $H$ in \equ{Ham} be real--analytic with $h$ steep in $\paramB$  with steepness indices $\parama_1$,...,$\parama_{n-1}$  and let
\begin{equation}
p_1:=\prod_{k=1}^{n-2}\parama_k\ ,\qquad
a:=\frac{1}{2n p_1}\ \ ,\qquad
b:={a\over \parama_{n-1}}\ .
\nonumber%\label{a.b}
\end{equation}
Then, there exist positive constants $\paramepsu,\paramo,T, c>0$
such that for any $0\leq \parameps<\paramepsu$ 
the solution $(I_t,\varphi_t)$
of  the   Hamilton equations 
for
 $H(I,\varphi)$ 
 with initial data $(I_0,\varphi_0)$ with $I_0\in U-2\paramo \parameps^b$
 satisfies 
\beq{main 1}
\Norm{I_t-I_0}\leq \paramo \parameps^b
\eeq
for any time $t$ satisfying:
\beq{main 2}
\norm{t}\leq {T\over \sqrt{\parameps}} 
\exp  \left ( {c\over \parameps^a}\right)
 .
\eeq
}

\vskip 0.4 truecm 
\noindent
{\it {\bf C}. Quantitative formulation.} Next, we provide explicit 
estimates for the parameters  $\paramepsu,\paramo,T, c$ appearing in Theorem~1.

\giu
To do this, we need to introduce some notations.
Given   
``extension parameters'' 
$\eta,\sigma>0$ and any set $D$,  
we let the ``extended complex domains'' be defined by:
$$
D_\eta= 
\bigcup_{I'\in D} 
\{ I\in {\mathbb C}^n:\ \ \Norm{I-I'}\leq \eta\}  \quad
{\rm 
and } \quad
{\mathbb T}^n_\sigma =\{
\varphi \in {\mathbb C}^n/(2\pi {\mathbb Z})^n:\ \ \norm{\Im \varphi_i}\leq \sigma \} .
$$
For any real action--angle function $u(I,\varphi)$ 
analytic in $D_\eta \times  {\mathbb  T}^n_\sigma$, 
with Fourier harmonics $u_k(I)$, we denote its Fourier--norm
$$
\norm{u}_{\eta,\sigma} =
\sum_{k\in {\mathbb Z}^n}\norm{u_k}_{{}_{D_\eta}}\,e^{|k| \sigma}\ ,
$$
where $\norm{\,.\,}_{D_\eta}$ denotes the sup--norm in $D_\eta$; if it needs to be specified, we shall also use the heavier notation $\norm{u}_{D;\eta,\sigma}$. 

\vskip 0.4 cm
\noindent
Let $H$ be real--analytic in $\paramB \times {\mathbb T}^n$ with $h$ steep in $\paramB$ with 
steepness indices 
 $\parama_1,\ldots ,\parama_{n-1}$ and  steepness coefficients
 $C_1,\ldots ,C_{n-1}$ and $\paramb$. 
Without loss of generality, we can take the extension parameter in action space  to be equal to the steepness coefficient $\paramb$
and
we can find
positive constants 
$\paramg,\paramom,\parame$ and $M$  such that:
\begin{itemize}
\item[$\bullet$] 
$h(I)$ is real analytic on an open set which contains $\paramB_{\paramb}$; 

\item[$\small \bullet$] 
$f(I,\varphi)$ is real analytic on an open set which contains $\paramB_{\paramb}\times {\mathbb
  T}^n_{\paramg }$; 

\item[$\small \bullet$] 
For any $I\in \paramB$,  we have:
$$
\paramom  \leq  \Norm{\omega(I)} \le \parame
$$
and, for any $I_1,I_2\in \paramB_{\paramb}$,  we have:
$$
\Norm{\omega(I_1)-\omega(I_2)}\leq M\Norm{I_1-I_2}  .
$$
\end{itemize}
Now, for $1\leq j\leq n-2$, let
\beq{par.aux}
p_j:=\prod_{k=j}^{n-2}\parama_k\ ,\quad\paramq_j:= 
n p_j-j \ , \quad \beta_j:=\parama_j+j(\parama_j-1)\ ,
\eeq 
and define the parameters 
\beqa{qj}
\paraml_j &:=&{\paramom \over M} \Big ({C_j \over 
\paramom}\Big )^{1\over \parama_j}+4 
\Big (2{2\parame +M\paramb \over \paramom}
\Big )^{1\over \parama_j}\ \ ,\\
E &:=& \max\left(  \max_{j\leq n-2}  
\left({(4M\paraml_j)^{\parama_j} \ 6^{\paramq_j(\parama_j-1)} \over C_j  
{\Big ({\paramom \over 2 \sqrt{2}}\Big )^{\parama_j-1}}} \right)^{\frac1{\beta_j}}\ ,\ 4 \right)  .
\label{econst}
\eeqa
Then, in Theorem~1, one can take
\begin{eqnarray}
\paramepso &:=& {1\over 2^8}{1\over 6^{4np_1-5}
E^{2np_1-1}}
{\paramom^2\over M\norm{f}_{\paramb,\paramg}}\cr
\paramepsu &:=& \paramepso
\ \min\left ( 
\Big ({6\sqrt{2}\over n}{M\paramb\over \paramom}\Big )^{1\over b} , \Big ({18\sqrt{2}\over n} \Big )^{ 1\over b} , 
\Big ({\paramb \over 4n \paraml_{n-1}}\Big )^{1\over b}
\Big ( {12\sqrt{2}E C_{n-1}\over \paramom}\Big )^{1\over a } , 
\Big ({\paramg  \over 6}\Big )^{ {1\over a}} , 1
\right )\cr
c&:=& \paramepso^a\  \frac{\paramg}{6}
\cr
\paramo &:=& \frac{\paramb\  n\ \mu_0}{\paramepso^b}\ ,\ {\rm with}\ \ \mu_0:= 
\max\left ( {1\over 24\sqrt{2}} {\paramom \over M\paramb},
{1 \over 6^2  2\sqrt{2}} , 
{\paraml_{n-1}\over \paramb} \Big ({\paramom \over 
12 \sqrt{2} E  C_{n-1}}\Big )^{1\over \parama_{n-1}} \right )
\cr
T&:=& {\paramg \over 24\sqrt{2}} 
{\paramom\over M (6E)^{1\over a} \sqrt{\paramepso}\norm{f}_{\paramb,\paramg} } .
\label{quantitative}
\end{eqnarray}

\vskip 0.4 truecm 
\noindent
{\it {\bf D}. On the proof.}  The proof of Nekhoroshev's theorem, 
in its various  settings, can be split into:
\\
 a {\it geometric part}, 
devoted to the analysis of distribution of small divisors in 
action--space;
\\
an {\it analytic part}, devoted to the construction of 
normal forms; 
\\
a {\it stability argument} yielding the confinement of the 
actions.\\
While the analytic part is obtained  by adapting averaging 
methods to an analytic setting, the heart of  Nekhoroshev's theorem 
resides in its geometric part. The geometric part of the steep case 
presented in \cite{Nekhoroshev, Nekhoroshev79} still needed a 
deep revisitation, which is performed here and  leads, in particular, to substantially improved    
stability exponents. 

\giu The proof of Theorem~1 will be obtained by deeply revisiting the 
geometric part of (\cite{Nekhoroshev, Nekhoroshev79}). The essential 
improvement are the following.
\\
First, 
we extend P\"oschel's Geometric Lemma (see \cite{poschel}) to allow for 
a more general power--law scaling of the amplitudes of the resonance domains. In  this way, 
we allow for a definition of the resonance domains which depends on the euclidean 
volume (of a minimal cell) of the lattice generating the resonance, 
and is compatible with steepness indices $\alpha_i>1$. In contrast with the 
convex case, the analog of P\"oschel's Geometric Lemma  
is here far to accomplish the geometric part of the theorem. In fact, 
motions with initial conditions characterized by a given resonance, 
may move along preferential planes of the action space, called fast drift 
planes. In particular, one needs to extend in the action space with fast drift planes
the resonant domains obtained by a pull back from the frequency space: eventual 
degeneracies of the frequency map, which are   typical of the steep non--convex case, may produce
topologically complicate sets.  
Nevertheless, a regularity of the distribution of these extended resonant sets  
must be proved: this is needed in order to  grant the non overlapping of resonant domains 
of the same multiplicity. In  \cite{Nekhoroshev,Nekhoroshev79}, the 
non--overlapping is granted simply by construction 
of the resonance domains, but the price paid was an overestimate
of resonant domains with the consequence of a strong $n^{-2}$ scaling 
of the stability exponent (\ref{expa}). Here, we do not grant the overlapping 
by construction but, with a careful analysis of the topology of these sets, we obtain 
a better balance between optimal definition of resonant domains and 
their non--overlapping. Finally, our geometric construction is fully 
compatible with the usual analytic part and stability argument, 
such as the so called {\it resonance trap} of \cite{Nekhoroshev,Nekhoroshev79}), and its improved version introduced 
in \cite{BGG}. 
\vskip 0.4 truecm 
\noindent
{\it {\bf E}.} The paper is organized as follows. The main part (i.e., the geometric analysis) is presented in \S~2:
in \S~2.1 we introduce several auxiliary parameters (needed to measure various covering sets, small divisors,
cut-offs in Fourier space, time scales, etc.) and point out the relevant relations among them (relations, which, although based on simple calculus, are proven, for completeness in Appendix B).  In \S~2.2, merging and extending the geometric 
analysis of \cite{Nekhoroshev} and \cite{poschel}, we introduce a covering in action--space formed by  (a suitable scale of) resonant and non--resonant  regions. Section~2.3 is the heart  of the paper, where the relevant analytic properties
of the resonant and non--resonant regions are proven; the section is divided into three lemmata:
the first is about geometric estimates concerning resonant domains; the second deals with small divisor estimates and the third one is a non--overlapping result for  resonant regions corresponding to resonances of the same dimension. In \S~3 we recall briefly P\"oschel's normal form theory\footnote{Incidentally, real--analyticity is needed only here; in the geometric analysis $C^2$--regularity is enough.} \cite{poschel} and show how it can be used in our setting. In the final 
section~4 we put all pieces together and prove Theorem~1 with the constants listed in {\bf C} above.  In Appendix~A we briefly review the notion of angles between linear spaces and, as mentioned above, Appendix~B is an elementary check of the main relations among the auxiliary parameters.

\vskip 0.4 truecm 
\noindent
{\footnotesize
{\bf Acknowledgement.}
We are indebted with L. Niederman for kindly providing Ref \cite{Nie.cor} prior to publication.
}

\section{Geometry of resonances}
\label{sec:geometry}

\subsection{Auxiliary parameters}
In the proof of the Theorem~1
several auxiliary parameters will occur; in this section we define such parameters and point out some
compatibility relations expressed as in inequalities, which will be needed in the following.

\begin{eqnarray}
&& K:= \Big ({\paramepso\over \parameps}\Big )^a
\\
&& 
\paramr :=2 \paramo\parameps^b\\
&&
\paramh := \frac{\paramr}{2n}\ ,
\\
&& \paramf :=  {\paramom \over 2\sqrt{2}}  \\
&&
\paramq_n:= 0\ ,\quad \paramq_{n-1}:=1\ ,\quad
a_{n-1}:=1\ ,\quad
a_j := \paramq_j-\paramq_{j+1}  \quad (1\le j\le n-2)
\ .
\label{a}
\end{eqnarray}
Notice that the $\paramq_j$'s are strictly decreasing since $a_j
\ge 1$, indeed:
\begin{equation}\label{aux.B.3}
a_{n-2}=n(\parama_{n-2}-1)+1\ ;\quad
a_j=np_{j+1}(\parama_j-1)+1\ , \  (1\le j\le n-3)
\ .
\end{equation}
Let $\Lambda$  be any maximal $K$--lattice 
over ${\mathbb Z}^n$ of 
dimension\footnote{We   recall that a ``maximal $K$--lattice'' $\Lambda$ is a lattice which  admits a basis of vectors $\tilde k\in  {\mathbb Z}^n$ with 
$\norm{\tilde k}:= \sum_{i=1}^n \norm{\tilde k_i}\leq K$, and it is not properly 
contained in any other lattice of the same dimension; the volume $|\Lambda|$
of the lattice $\Lambda$ is defined as the euclidean volume of the parallelepiped spanned by a basis for $\Lambda$; (see \cite{poschel}). Notice that for any $K$--lattice of dimension $j$, one has  $1\le |\Lambda|\le K^j$.}
$1\le j \le n-1$, 
$|\Lambda | $ its volume,  and set:
\begin{eqnarray}
&&\lambda_j :=  {\paramf\over (AK)^{\paramq_j}}
 \ , \qquad {\rm where} \  A:=6E
 \label{condlambda2}\\
&&
r_j := \paraml_j \Big ({\lambda_j\over C_j}\Big )^{1\over \parama_j}  \qquad\label{rj}\\
&&\paramp_\Lambda:={\lambda_j\over \norm{\Lambda}}\  
\\
&&\paramk_\Lambda :={\paramp_\Lambda\over M}  \label{rlam}
\\
&&\paramc_\Lambda := (E K)^{a_j}\paramp_\Lambda  
\label{alphalambda}\\
&&\paramd_\Lambda:={\paramc_\Lambda\over 4MK}  ,\label{rhoi}
\end{eqnarray}
Finally, we set
\begin{eqnarray}
&&\paramn:= {\lambda_1\over 2MK}
\\
&&T_0:={\paramg  \paramn \over 5 \parameps
\norm{f}_{\paramb,\paramg}}
e^{K {\paramg \over 6}} \ ,\qquad
T_\Lambda := {e\,\paramg  \over 24}
{\paramk_\Lambda\over \parameps\norm{f}_{\paramb,\paramg} }e^{K {\paramg \over 6}}
\ ,\quad 
T_j := \min_{\Lambda: {\rm dim}\Lambda=j}T_{\Lambda}   
\label{TT}
\\
&&  T_{\rm exp}:= \min_{i=0,\ldots,n-1}T_i  .
\label{TTT}
\end{eqnarray}
It is then easy to check  (see Appendix \ref{inequalities})  that under the assumption of 
Theorem~1, namely,  $0\le \parameps< \paramepsu$, for any maximal $K$--lattice of dimension $1\le j\le n-1$ (unless otherwise specified)
one has:
\begin{eqnarray}
&& A \geq \max\left( \max_{j\in \{1,\ldots ,n-1\}}\Big ((E^{a_j}+1)^2+1\Big )^{1\over 2 a_j} \ ,\
\Big ({ 4\over E^{a_j}}+2\Big )^{1\over a_j}\right)
\label{conda1}
\\
&&
K\paramg\ge 6
\label{Ksigma}
\\
\label{conditionsin.1}
&& \paramk_\Lambda \leq  \min \Big ({\paramh\over 2}, \paramd_\Lambda\Big )
 \ ,
 %\qquad  (j\le n-2)
\\
&& 
\paramp_\Lambda \leq \min \Big 
(  {\paramom\over \paramb } {\paramh\over 4} \ ,\ 
{\paramom\over  2\paramb} (\paramh-\paramk_\Lambda)\ ,\   
\paramf \Big ) , \qquad (j\le n-2)
\label{conditionsin.2}
\\
&&
KM\paraml_j\Big ({\paramp_\Lambda\over C_j}\Big )^{1\over \parama_j}
\leq {1\over 4}\paramc_\Lambda \ ,\qquad  (j\le n-2)
\label{conditionsin.3}
\\&&
\label{rhoLrho}
\paramd_\Lambda\le \paramb\ ,
\\
&&
\label{epsnonres}
\parameps \norm{f}_{\paramb,\paramg}  
\leq 
\min\left(
{1\over 2^8}{\lambda_1 \paramn\over K}
\ ,\  {\paramc_\Lambda \paramk_\Lambda\over 2^9 K} 
\ ,\  {\paramc_\Lambda \paramd_\Lambda\over 2^9 K} \right)
\\
\label{rho est}
&&\max_{0\le i\le n-1} \ r_i  \le \paramh
\end{eqnarray}
\begin{eqnarray}
&&
\label{rho0}
\paramn\le \paramb 
\\
&& \label{rho0r}
\sum_{j=0}^{n-1} r_j\le {\paramr\over 2}
\\
&& 
\label{rleqrho}
\paramr \leq \frac\paramb2  \ ,
\\
&& 
\label{Texp}
T_{\rm exp}\ge {T\over \sqrt{\parameps}} \exp\Big({K \paramg\over 6}\Big)\ .
\end{eqnarray}
 
\subsection{Resonant and non--resonant domains} 

Fix $I_0\in
\paramB-\paramr$ and consider the set:
$$
\paramD := \B(I_0,\paramr) \subseteq \paramB  .
$$
In order to prove the stability of all motions with initial actions $I_0$, 
we need to cover the domain $\paramD$ with open domains 
where suitable normal forms adapted to the local resonance properties 
may be constructed. We here introduce resonant zones and resonant blocks  
as in \cite{poschel}, but, since  we do not require any local inversion for the 
frequency map $\omega(I)$ (as it is typical of steepness 
\cite{Nekhoroshev,Nekhoroshev79}, see also \cite{GLF11}), 
these domains are directly defined in the action--space, without using 
any pull--back from a frequencies space. Then, we define suitable extensions, 
in the spirit of the original construction of \cite{Nekhoroshev} 
(see also \cite{BGG}). 
 
\noindent 
We first define the resonant zones and blocks depending on the 
parameter $K\geq 1$, representing a cut--off for the resonance order, and also on the parameters $0<\lambda_1<\ldots <\lambda_{n-1}<\paramf$ defined above.
As in \cite{poschel}, we  
consider   only the resonances defined by
$$
k\cdot \omega(I)=0
$$
with  $k$ in some maximal $K$--lattice 
$\Lambda\subseteq {\mathbb Z}^n$. We define the {\bf resonant 
zone}
\begin{equation}
{\mathcal Z}_\Lambda := \{ I\in \paramD: \ \ \Norm{\pi_{\langle\Lambda\rangle} \omega(I)}< 
\paramp_\Lambda\} ,
\end{equation}
where $\langle\Lambda\rangle$ denotes the real vector space spanned 
by the lattice $\Lambda$, and the {\bf resonant block}
\begin{equation}
\paramD_\Lambda := {\mathcal Z}_\Lambda \backslash {\mathcal Z}_{j+1}\ \ ,\ \ j=\dim \Lambda  ,
\end{equation}
where:
$$
{\mathcal Z}_{i}:=\cup_{\{\Lambda':\ \ \dim \Lambda'=i\}}{\mathcal Z}_{\Lambda'}  .
$$
We also define   ${\mathcal Z}_0:=\paramD$ and the {\bf non--resonant block} 
$\paramD_0$ by
$$
 \paramD_0:={\mathcal Z}_0
\backslash {\mathcal Z}_1  .
$$
We remark that, since $\Norm{\omega(I)}\ge \paramom>\paramf\ge \paramp_\Lambda$ for any $I\in \paramD$, 
the completely resonant zone ${\mathcal Z}_{{\mathbb Z}^n}$ is empty and so is ${\mathcal Z}_n$. This 
implies 
\beq{Dlam}
\paramD_\Lambda={\mathcal Z}_\Lambda\ ,\qquad \forall\   \Lambda\ {\rm s.t.}\quad 
\dim \Lambda= n-1\ .
\eeq
Furthermore, if one defines
$$
\paramD_j:=\cup_{\{\Lambda': \dim \Lambda' =j\}} \paramD_{\Lambda'} \ ,
$$
one sees immediately that
$$
\paramD_j= {\mathcal Z}_j \setminus {\mathcal Z}_{j+1}  ,
$$
so that, for any $1\leq j\leq n-1$, we have:
\beq{Dj}
\paramD= \paramD_0 \cup \paramD_1 \cup ... \cup \paramD_{j-1} \cup {\mathcal Z}_j  ,
\eeq
and, in particular,
\beq{covering}
\paramD= \paramD_0 \cup \paramD_1 \cup ... \cup \paramD_{n-1}\ .
\eeq
Next, following Nekhoroshev, we introduce {\bf discs}
\begin{equation}\label{cylinders}
{\discs}_{\Lambda,\eta}^\paramh(I):=\left(\left(\bigcup_{I'\in I+\langle\Lambda\rangle} 
\B(I',\eta)\right) \cap 
{\mathcal Z}_\Lambda \cap (\paramD-\paramh)\right )^I   
\subseteq 
{\mathcal Z}_\Lambda \cap (\paramD-\paramh),
\end{equation}
where $I+\langle\Lambda\rangle $ (called by Nekhoroshev, ``fast drift plane'') denotes the plane through $I$ parallel to 
$\langle\Lambda\rangle$,
$\left(C \right)^I$ denotes the connected component of a set $C$ which 
contains $I$ and $\eta$ is any positive number less or equal than $\paramh$.
The  {\bf extended resonant blocks} are then defined by\footnote{Notice that, if $I\in \paramD_\Lambda$, then $I\in {\discs}_{\Lambda,\eta}^\paramh(I)$ so that 
$\paramD_\Lambda\cap (\paramD-\paramh)\subset \paramD_{\Lambda ,\paramk_\Lambda}^\paramh$.\label{ftn:inc} }:
\begin{equation}\label{paramD}
\paramD_{\Lambda ,\paramk_\Lambda}^\paramh:=
\bigcup_{I\in \paramD_\Lambda\cap (\paramD-\paramh) }{\discs}_{\Lambda ,\paramk_\Lambda}^\paramh(I)
\subset {\mathcal Z}_\Lambda\cap (\paramD-\paramh)
 ,
\end{equation}
and the {\bf extended non--resonant block} by:
$$
\paramD_0^\paramh:=\paramD_0 \cap (\paramD-\paramh).
$$
We remark that the set $\paramD-\paramh$ is not empty  since $\paramh <\paramr$,  
and for any lattice $\Lambda$ with $\dim \Lambda=n-1$, we have, by \equ{Dlam}, \equ{paramD} and footnote~\ref{ftn:inc}, 
\begin{equation}\label{paramDn}
\paramD_{\Lambda ,\paramk_\Lambda}^\paramh=\paramD_\Lambda \cap (\paramD-\paramh)\ ,\qquad(\dim \Lambda=n-1)\ .
\end{equation}

\subsection{Geometric properties of the  resonant domains}

\noindent $\bullet$ \underline{\sl Geometric estimates for resonant domains}

\giu
For any maximal $K$--lattice $\Lambda$, we 
need to estimate the diameter of the intersection of the fast drift 
planes $I+\langle\Lambda\rangle$ with the resonant zones:

\begin{lemma} \label{lem:actions}
For any $I'\in \paramD_\Lambda\cap(\paramD-\paramh)$ and $I''\in {\cal C}^\paramh_{\Lambda, \paramk_\Lambda}(I')$ we have: 
\begin{equation}\label{II''}
\Norm{I'-I''}\leq  \paraml_j 
\Big ({\paramp_\Lambda \over C_j} \Big )^{1\over \parama_j} \le r_j\ .
\end{equation}
\end{lemma}

\proof We divide the proof of this lemma in three steps.

\noindent
{\bf Step 1.} Let  $\tilde \paramp,\tilde\paramh>0$ be such that  
\begin{equation}
\tilde \paramp  \leq \min \Big ({{\tilde\paramh} \over \paramb}, 
{1\over \sqrt{2}}\Big )\paramom\ \ ,
\label{sigmaldelta}
\end{equation}
and
define 
\begin{equation}
{\mathcal Z}_\Lambda ({\tilde \paramp} ) = 
\{ I \in \paramD:\ \ \Norm{\pi_{\langle\Lambda\rangle} \omega(I)} < {\tilde \paramp} \}\ \ .
\end{equation}
Let us also denote by $\lomegar$ the linear space generated 
by $\omega(I)$; by $\fdp$ the linear space orthogonal to $\omega(I)$ and by 
$\lambdap = \pi_{\fdp} \langle\Lambda\rangle$ 
the linear space obtained by projecting every vector $u$ of 
$\langle\Lambda\rangle$ on $\fdp$. 

\noindent
The first step will consist in proving that:

\noindent
{\sl  For any  $I\in 
{\mathcal Z}_\Lambda ({\tilde \paramp} )\cap (\paramD-\tilde \paramh)$ and any
$I'\in \Big ((I+\langle\Lambda\rangle)\cap {\mathcal Z}_\Lambda({\tilde \paramp} )
\cap (\paramD-{\tilde\paramh})\Big )^I$ one has}: 
\beq{L: diam1}
\Norm{I-I'} <  4\Big ({2\parame +M\paramb \over \paramom}\,
{{\tilde \paramp}  \over C_j}\Big )^{1\over \parama_j}\ \ .
\eeq
Fix $I'\in 
\left ((I+\langle\Lambda\rangle )\cap 
{\mathcal Z}_\Lambda({\tilde \paramp} )\cap (\paramD-{\tilde\paramh})\right )^I$, with $I'\ne I$ (if 
$I'=I$ there is nothing to prove).  Then, there exists a curve\footnote{Notice that the set $\left ((I+\langle\Lambda\rangle )\cap 
{\mathcal Z}_\Lambda({\tilde \paramp} )\cap (\paramD-{\tilde\paramh})\right )^I$
is open in the relative topology of $I+\langle \Lambda\rangle$ and therefore is arc--connected in 
$I+\langle \Lambda\rangle$.
} 
$u(\params)\in \langle\Lambda\rangle$ such that $u(0)=0$, $u(1)=I'-I$,   
and for any $\params$,  
$I+u(\params)\in \left (\Big (I+\langle\Lambda\rangle\Big )\cap
{\mathcal Z}_\Lambda({\tilde \paramp} )
\cap (\paramD-{\tilde\paramh})\right )^I$. 
In particular,
$ \Norm{\pi_{\langle\Lambda\rangle} \omega(I+u(\params))} < {\tilde \paramp}$.

\giu
The proof of \equ{L: diam1} will be   based on the following claims (i)$\div$(vii).

\giu
(i)  {\sl $\lambdap$ is a vector space of dimension $j$}.

 \noindent{\sl Proof of} (i): 
Clearly, if $u_1,\ldots ,u_j$ is a basis for $\langle\Lambda\rangle$, 
then any vector in $\lambdap$ can be represented as a linear combination 
of $\pi_{\fdp}u_1$, ..., $\pi_{\fdp}u_j\in \lambdap$. We prove that the 
vectors  $\pi_{\fdp}u_1, \ldots ,\pi_{\fdp}u_j\in \lambdap$ are linearly 
independent, so that $\dim\lambdap=j$.
First, we remark that the only vector $u$ of 
$\langle\Lambda\rangle$ satisfying: $\pi_{\fdp}u=0$ 
is $u=0$. In fact, if there exists $u\ne 0$ such that 
$u\in  \langle\Lambda\rangle$ and $\pi_{\fdp}u=0$, then 
$\omega(I) \in \langle\Lambda\rangle$, and therefore we have:
$$
 \Norm{\omega(I)}=\Norm{\pi_{\langle\Lambda\rangle} \omega(I)}<  
{\tilde \paramp}  \leq {\paramom\over \sqrt{2}} ,
$$
which is not possible since for any 
$I\in \paramD$ we assumed $\Norm{\omega(I)} > \paramom$. 
Now, let us consider $c_1,\ldots ,c_j$ such that:
$\sum_{i=1}^j c_i \pi_{\fdp}u_i =0$. Then,  
$\pi_{\fdp}\sum_i c_i u_i =0$, and therefore $ \sum_i c_i u_i=0$. 
But, since the $u_i$ are linearly 
independent, it follows $c_1,\ldots ,c_j=0$.

\giu
(ii) {\sl For any $u\in \langle\Lambda\rangle$, we have
$\dst 
\pi_{\lambdap}u = \pi_{\fdp}u$.}

 \noindent{\sl Proof of} (ii): 
We first compute:
\begin{equation}\label{bbcc}
\pi_{\fdp}u = \pi_{\lambdap} \pi_{\fdp}u+
\pi_{{\lambdap}^{\perp}} \pi_{\fdp}u \ \ .
\end{equation}
Since $\pi_{\fdp}u \in {\lambdap}$, we have 
$\pi_{{\lambdap}^{\perp}} \pi_{\fdp}u=0$, so that \equ{bbcc} 
becomes:
\begin{equation}
\pi_{\fdp}u =  \pi_{\lambdap} \pi_{\fdp}u \ \ .
\label{equala}
\end{equation}
But, $\pi_{\lambdap} u = \pi_{\lambdap} ( \pi_{\fdp}u +
\pi_{\lomegar}u)= \pi_{\lambdap}\pi_{\fdp}u+
\pi_{\lambdap}\pi_{\lomegar}u$ and 
since $\lambdap \subseteq \fdp$, we  have 
$\pi_{\lambdap}\pi_{\lomegar}u=0$, and therefore:
\begin{equation}
\pi_{\lambdap} u = \pi_{\lambdap}\pi_{\fdp}u  .
\label{equal}
\end{equation}
>From equations (\ref{equala}) and (\ref{equal}) we get (ii).

 \giu
(iii) {\sl The angle\footnote{The notion of angle between linear spaces is briefly reviewed in Appendix~\ref{appendix angles}.}  
between $\langle\Lambda\rangle$ and $\lambdap$  is equal to the angle  between $\omega(I)$ and ${\langle\Lambda\rangle}^{\perp}$, in formulae:}
\begin{equation}
\langle\Lambda\rangle \angolo \lambdap = \omega(I)  \angolo  \langle\Lambda\rangle^{\perp} .
\label{altraest}
\end{equation}
 \noindent{\sl Proof of} (iii): By (ii) we have:
$\dst \langle\Lambda\rangle \angolo \lambdap = \max_{u\in \langle\Lambda\rangle, u \ne 0} 
u \angolo \pi_{\lambdap}u =
\max_{u\in \langle\Lambda\rangle, u \ne 0} 
u \angolo \pi_{\fdp}u$ $=$   
$\langle\Lambda\rangle\angolo \fdp$, 
and using (x) of Appendix~\ref{appendix angles}, we obtain $\langle\Lambda\rangle \angolo \lambdap 
=  \langle\Lambda\rangle\angolo {\fdp}=
\lomegar \angolo \langle\Lambda\rangle^{\perp}$.

\giu
(iv) {\sl For any $\params$, one has}
$\dst 
\Norm{\pi_{\lambdap} \omega(I+u(\params))}<
\frac{2\, \parame \,  \tilde \paramp}{\paramom}$.

 \noindent{\sl Proof of} (iv): 
We start with
\begin{eqnarray*}
\Norm{\pi_{\lambdap} \omega(I+u(\params))} &=&\sqrt{\Norm{\omega(I+u(\params))}^2-
\Norm{\pi_{\lambdap^\perp} \omega(I+u(\params))}^2}\cr
&=&  \Norm{\omega(I+u(\params))} \sqrt{ 1 - 
\norm{\cos(\omega(I+u(\params)) \angolo \lambdap^\perp)}^2}\cr
&=&  \Norm{\omega(I+u(\params))} \norm{\sin(\omega(I+u(\params)) \angolo \lambdap^\perp)}\cr
&\leq& {\parame}\norm{\sin(\omega(I+u(\params)) \angolo \lambdap^\perp)}
\end{eqnarray*}
and then we produce an upper bound estimate of the angle 
$\omega(I+u(\params)) \angolo \lambdap^{\perp}$.  By using property (ix) of  Appendix~\ref{appendix angles}, we 
first obtain:
\begin{equation}
\omega(I+u(\params)) \angolo \lambdap^{\perp} \leq 
\omega(I+u(\params)) \angolo \langle\Lambda\rangle^{\perp}+\langle\Lambda\rangle^{\perp} \angolo 
\lambdap^{\perp} .
\label{angolo}
\end{equation}
Now, recalling that  $\langle\Lambda\rangle$ and $\lambdap$ have the
same dimension (claim (i) above), we see that  by properties (x) and (xi) of Appendix~\ref{appendix angles},  
$\dst 
\langle\Lambda\rangle^{\perp} \angolo \lambdap^{\perp}= \lambdap \angolo 
\langle\Lambda\rangle= \langle\Lambda\rangle \angolo \lambdap=
\omega(I)\angolo \langle\Lambda\rangle^{\perp}$.
>From (\ref{angolo}), we therefore obtain:  
\begin{equation}
\omega(I+u(\params)) \angolo \lambdap^{\perp} \leq  
\omega(I+u(\params)) \angolo \langle\Lambda\rangle^{\perp}+
\omega(I)\angolo \langle\Lambda\rangle^{\perp} \ \ .
\label{twoangles}
\end{equation}
Then, since:
\begin{eqnarray}
&&\norm{\sin \Big(\omega(I+u(\params)) \angolo \langle\Lambda\rangle^{\perp}\Big)} = 
{\Norm{\pi_{\langle\Lambda\rangle} \omega(I+u(\params))}\over 
\Norm{\omega(I+u(\params))}} < {{\tilde \paramp}  \over \paramom}
\cr
&& \norm{\sin \Big(\omega(I) \angolo \langle\Lambda\rangle^{\perp}\Big)} = 
{\Norm{\pi_{\langle\Lambda\rangle} \omega(I)}\over 
\Norm{\omega(I)}} < {{\tilde \paramp}  \over \paramom}
 ,
\end{eqnarray}
and ${\tilde \paramp}/ \paramom \leq 1/\sqrt{2}$, 
both angles are strictly smaller than $\pi/4$, 
their sum is strictly smaller than $\pi/2$, and since $\sin(x)$ is monotone 
in $[0,\pi/2]$, from (\ref{twoangles}) and  standard trigonometry,
we obtain:
\begin{eqnarray*}
\norm{\sin \omega(I+u(\params)) \angolo \lambdap^{\perp}}&\leq & 
\norm{\sin (\omega(I+u(\params)) \angolo \langle\Lambda\rangle^{\perp}+\omega(I)\angolo 
\langle\Lambda\rangle^{\perp})}\cr
&\leq& 
\norm{\sin(\omega(I+u(\params)) \angolo \langle\Lambda\rangle^{\perp})}+
\norm{\sin(\omega(I)\angolo \langle\Lambda\rangle^{\perp})} < 
2\, {{\tilde \paramp} \over \paramom} .
\end{eqnarray*}
We therefore obtain:
$\dst 
\Norm{\pi_{\lambdap} \omega(I+u(\params))}\leq 
{\parame} \norm{\sin \omega(I+u(\params)) \angolo \lambdap^{\perp}}
< \frac{ 2\, \parame\, \tilde \paramp}{\paramom}$.

\giu
(v) $\dst  \Norm{\pi_{\lomegar}u(\params)} < {{\tilde \paramp} \over \paramom}
\Norm{u(\params)}$.

\giu
{\sl Proof of} (v):
Since $u(\params)\in \langle\Lambda\rangle$ and 
$I\in {\mathcal Z}_\Lambda ({\tilde \paramp} )$, we have:
$$
\Norm{\pi_{\lomegar}u(\params)} =
{ \norm{\omega(I)\cdot u(\params)}\over \Norm{\omega(I)}} 
={\norm{\pi_{\langle\Lambda\rangle} \omega(I) \cdot u(\params)}\over 
\Norm{\omega(I)}} < {{\tilde \paramp} \over \paramom} \Norm{u(\params)}\ \ .
$$
(vi) $\dst  I+\pi_{\fdp}u(\params) \in \paramD$.

\noindent
{\sl Proof of} (vi):
Since $I,I+u(\params)\in \paramD-{\tilde\paramh}$, we have
$\Norm{u(\params)}\leq 2 r$ and, using (\ref{rleqrho}), we obtain
$\Norm{u(\params)}\leq \paramb$. Then, from (v) and (\ref{sigmaldelta}), 
we have:$\dst \Norm{\pi_{\lomegar}u(\params)} <  {{\tilde \paramp}  \over \paramom}
\Norm{u(\params)} \leq  {{\tilde \paramp}  \over \paramom}\paramb\leq {\tilde\paramh}$.
Therefore, $I+\pi_{\fdp}u(\params) \in \paramD$.

\giu
(vii)  $\xi:=\|\pi_{\fdp}(I'-I)\| \in (0,\paramb ]$. 

\noindent{\sl Proof of} (vii): 
Let us first 
assume $\xi=0$, that is $I'-I\in \lomegar$ so that
$$
I'-I = \omega(I) {\Norm{I'-I}\over \Norm{\omega(I)}} .
$$
Since $I'-I\in \langle\Lambda\rangle$ and $I'\ne I$, this would imply also
$\omega(I)\in \langle\Lambda\rangle$, and therefore: 
$$
 \Norm{\omega(I)}=\Norm{\pi_{\langle\Lambda\rangle} \omega(I)}<  
{\tilde \paramp}  \leq {\paramom\over \sqrt{2}}  ,
$$
which is not possible since for any 
$I\in \paramD$ we have $\Norm{\omega(I)} > \paramom$. Therefore we have $\xi>0$. 
Then, we have 
$$
\xi =\Norm{\pi_{\fdp}(I'-I)} \leq 
\Norm{I'-I}  = \Norm{u(1)}  \leq \paramb  .
$$
Now, we are ready to complete the proof of \equ{L: diam1}.
Since $0< \xi \leq \paramb$, let $0\leq \eta_*\leq \xi$ the $\eta$ 
which realizes the maximum in the definition  of the steepness index of
dimension $j$, that is:
\begin{equation}
\min_{u\in {\lambdap}:\ \Norm{u}=\eta_*}
\Norm{\pi_{\lambdap} \omega (I+u)} > C_j \xi^{\parama_j}  .
\label{eqlem}
\end{equation}
The curve $\pi_{\fdp}u(\params)\in \lambdap$ joins $I$ and $I+\pi_{\fdp}(I'-I)$, 
and therefore 
$$[0,\xi ] \subseteq \cup_{\params\in [0,1]} \Norm{\pi_{\fdp}u(\params)}\ ,$$ 
so that there exists $\params_*\in [0,1]$ such that 
$\Norm{\pi_{\fdp}u(\params_*)}= \eta_*$. From (\ref{eqlem}) it follows:
$$
\Norm{\pi_{\lambdap} \omega (I+\pi_{\fdp}u(\params_*)} > C_j 
\xi^{\parama_j}  .
$$
But using claims (iv) and (v)  we also obtain:
\beqano
\Norm{\pi_{\lambdap} \omega (I+\pi_{\fdp}u(\params_*)} &\leq& 
\Norm{\pi_{\lambdap} \omega (I+u(\params_*))} + 
M \Norm{\pi_{\lomegar}u(\params_*)}<
2{\parame\over \paramom}{\tilde \paramp} 
+M {{\tilde \paramp}  \over \paramom}
\Norm{u(\params_*)}  \\
&<& 
{2\parame +M\paramb  \over \paramom}\, {\tilde \paramp}  
\eeqano
so that
$$
C_j \xi^{\parama_j} <  
{2\parame +M\paramb  \over \paramom}\,{\tilde \paramp}  ,
$$
and therefore
$$
\Norm{\pi_{\fdp}(I'-I)}=\xi  < 
\Big ( {2\parame +M\paramb  \over \paramom}\, {{\tilde \paramp}  \over C_j}
\Big )^{1\over \parama_j} .
$$
Using again (v), we obtain:
\begin{eqnarray}
\Norm{I'-I}&\leq& 
\Norm{\pi_{\fdp}(I'-I)}+\Norm{\pi_{\lomegar}(I'-I)}
\cr
&< &\Big ({2\parame +M\paramb  \over \paramom}
{{\tilde \paramp}  \over C_j}\Big )^{1\over \parama_j} +
{ {\tilde \paramp} \over \paramom} \Norm{I'-I}\cr
&\leq& \Big ({2\parame +M\paramb  \over \paramom}
{{\tilde \paramp} \over C_j}\Big )^{1\over \parama_j} +
{1\over \sqrt{2}}\Norm{I'-I} ,
\end{eqnarray}
that is:
$$
\Norm{I'-I} <  {1\over 1-{1\over \sqrt{2}}} 
\Big ({2\parame +M\paramb  \over \paramom}
{ {\tilde \paramp} \over C_j}\Big )^{1\over \parama_j} <  
4\, \Big ({2\parame +M\paramb  \over \paramom}
{ {\tilde \paramp} \over C_j}\Big )^{1\over \parama_j} .
$$
This finishes the proof of \equ{L: diam1}.

\vskip0.5cm
\noindent
{\bf Step 2.} Next, we prove that:

 \noindent {\sl For any 
$I\in {\mathcal Z}_\Lambda \cap (\paramD-\paramh)$ and any
 $I'\in {\discs}_{\Lambda,{\paramk_\Lambda}}^\paramh(I)$, we have:
\begin{equation}
\Norm{I-I'} \leq {\paramk_\Lambda}  + 4 \Big ({2\parame +M\paramb  \over \paramom}
{\paramp_\Lambda+M{\paramk_\Lambda} \over C_j}\Big )^{1\over \parama_j} .
\label{diamcilindr}
\end{equation}
}

\noindent
Fix $I'\in {\discs}_{\Lambda,{\paramk_\Lambda}}^\paramh(I)$. Since 
${\discs}_{\Lambda,{\paramk_\Lambda}}^\paramh(I)$ is open and connected, there exists a 
curve $I+u(\params)\in {\discs}_{\Lambda,{\paramk_\Lambda}}^\paramh(I)$, $\params\in [0,1]$,  
such that $I+r(0)=I$, $I+u(1)=I'$. 
Since 
${\discs}_{\Lambda,{\paramk_\Lambda}}^\paramh(I)\subseteq {\mathcal Z}_\Lambda$, we have:
$\Norm{\pi_{\langle\Lambda\rangle} \omega(I+u(\params))}< \paramp_\Lambda
$ for any $\params\in [0,1]$, and also
$\Norm{\pi_{{\langle\Lambda\rangle}^\perp} u(\params)}\leq {\paramk_\Lambda}$.
In fact, since $I+u(\params)\in {\discs}_{\Lambda,{\paramk_\Lambda}}^\paramh(I)\subseteq 
\cup_{\tilde I\in I+\langle\Lambda\rangle}\B(\tilde I,{\paramk_\Lambda})$, 
there exists a curve $u'(\params)\in \langle\Lambda\rangle$ such that 
$\Norm{u(\params)-u'(\params)}\leq {\paramk_\Lambda}$, and therefore 
$$\Norm{\pi_{{\langle\Lambda\rangle}^\perp} u(\params)}= 
\Norm{\pi_{{\langle\Lambda\rangle}^\perp} (u(\params)-u'(\params)) }\leq 
\Norm{u(\params)-u'(\params)}\leq {\paramk_\Lambda}\ .
$$ 
Then, we define $u''(\params):=\pi_{\langle\Lambda\rangle} u(\params)$, so that 
$ u''(0)=\pi_{\langle\Lambda\rangle} u(0)=0$, 
$I+u''(\params)\in I+\langle\Lambda\rangle$, and 
$$
\Norm{u''(\params)-u(\params)}=\Norm{\pi_{\langle\Lambda\rangle}u(\params)-u(\params)}
=\Norm{\pi_{{\langle\Lambda\rangle}^\perp}u(\params)}\leq {\paramk_\Lambda}  .
$$
Therefore, on the one hand we have $I+u''(\params)\in \paramD-\paramh+{\paramk_\Lambda}$, on the 
other hand:
\begin{eqnarray*}
\Norm{ \pi_{\langle\Lambda\rangle} \omega(I+u''(\params))}&\leq&  
\Norm{ \pi_{\langle\Lambda\rangle} \omega(I+u(\params))}+
\Norm{ \pi_{\langle\Lambda\rangle}  (\omega(I+u''(\params))-\omega(I+u(\params)))}\\
&\leq& \Norm{ \pi_{\langle\Lambda\rangle} \omega(I+u(\params))}+
\Norm{\omega(I+u''(\params))-\omega(I+u(\params))}\\
&\leq& \Norm{ \pi_{\langle\Lambda\rangle} \omega(I+u(\params))}+M\Norm{u''(\params)-u(\params)}\leq 
\paramp_\Lambda+ M{\paramk_\Lambda}  .
\end{eqnarray*}
Therefore, for any $t\in [0,1]$, we have:
$$
I+u''(\params)\in  \Big ((I+\langle\Lambda\rangle)\cap {\mathcal Z}_\Lambda(\paramp_\Lambda+M{\paramk_\Lambda})\cap (\paramD-(\paramh-{\paramk_\Lambda}))\Big )^I .
$$
We use, now, \equ{L: diam1}  (step 1) 
with 
$$\tilde \paramp:= \paramp_\Lambda+M{\paramk_\Lambda}\ \qquad {\rm 
and} \qquad
\tilde \paramh:=\paramh-{\paramk_\Lambda}\ .
$$
In fact, $I\in {\mathcal Z}_\Lambda\subseteq 
 {\mathcal Z}_\Lambda(\tilde \paramp )$; 
$I \in \paramD-\paramh \subseteq \paramD-  \tilde \paramh$; from \equ{conditionsin.2}
it follows:
$$
 \tilde \paramp  \leq \min \Big 
( {\paramom\over \paramb}\tilde \paramh \ ,\   
{\paramom\over \sqrt{2}}\Big ) .
$$
Therefore, we have:
$$
\Norm{u''(\params)}\leq 4 \Big ({2\parame +M\paramb  \over \paramom}\,
{\tilde \paramp  \over C_j}\Big )^{1\over \parama_j}=
4 \Big ({2\parame +M\paramb  \over \paramom}\,
{\paramp_\Lambda+M{\paramk_\Lambda}  \over C_j}\Big )^{1\over \parama_j} ,
$$
for any $\params\in[0,1]$. In particular we have:
$$
\Norm{I'-I}=\Norm{u(1)}\leq \Norm{u''(1)-u(1)}+\Norm{u''(1)}
\leq {\paramk_\Lambda}  + 4 \Big ({2\parame +M\paramb  \over \paramom}
{\paramp_\Lambda+M{\paramk_\Lambda}  \over C_j}\Big )^{1\over \parama_j} .
$$
{\bf Step 3.} We may conclude the proof of the lemma.
>From (\ref{conditionsin.1}) and 
(\ref{conditionsin.2}) we obtain 
$$
\paramp_\Lambda+M \paramk_\Lambda =  2 \paramp_\Lambda \leq {\paramom \paramh \over 
2\paramb} \leq {\paramom \over \paramb}(\paramh- \paramk_\Lambda)  ,
$$
so that applying \equ{diamcilindr} 
and using again (\ref{conditionsin.1}), 
we have:
$$
\Norm{I'-I''}\leq  \paramk_\Lambda+ 4 \Big ({2\parame +M\paramb  \over \paramom}\,
{\paramp_\Lambda+M\paramk_\Lambda\over C_j}\Big )^{1\over \parama_j}
\leq {\paramp_\Lambda\over M}+ 
4 \Big (2{2\parame +M\paramb  \over \paramom}\, 
{\paramp_\Lambda\over C_j}\Big )^{1\over \parama_j} .
$$
Then, since $\parama_j \geq 1$ and (recall \equ{conditionsin.2}) $\paramp_\Lambda/\paramom <1$, we have 
$({\paramp_\Lambda / \paramom}) \leq  ({\paramp_\Lambda / \paramom}
)^{1\over \parama_j}$, from which the first inequality in \equ{II''} follows at once; the second inequality follows 
from the fact that $\paramp_\Lambda\le \lambda_j$ and from the definition of $r_j$.
\hfill $\Box$

\vskip0.5cm
\noindent
$\bullet$ \underline{\sl Small divisor estimates}

\giu
We recall (\cite{poschel}) that a set $\tilde \paramD\subseteq \paramD$ is 
$\paramc $--$K$ non resonant modulo $\Lambda$
if we have $\norm{k\cdot \omega(I)}\geq \paramc $ for any $k\in {\mathbb Z}^n\backslash \Lambda$ 
such that $\norm{k}\leq K$; we will say that $\tilde \paramD\subseteq \paramD$ is $\paramc$--non resonant if
 $\norm{k\cdot \omega(I)}\geq \paramc$ for any $k\in {\mathbb Z}^n\backslash \{0\}$ 
such that $\norm{k}\leq K$. The following result is a generalization of the Geometric Lemma in \cite{poschel}.

\begin{lemma} \label{L:geoml} 
{\rm (i)} For any maximal $K$--lattice $\Lambda$, the resonant block $\paramD_\Lambda$ 
is $\paramc_\Lambda$--$K$ non resonant modulo $\Lambda$,
while the non resonant block $\paramD_0$ is $\lambda_1$--$K$ non resonant.
\\
{\rm (ii)} If $j= n-1$,  the extended block $\paramD_{\Lambda,\paramk_\Lambda}^\paramh$ 
is $\paramc_\Lambda$--$K$ non--resonant modulo $\Lambda$; if $j\le n-2$, the extended block $\paramD_{\Lambda,\paramk_\Lambda}^\paramh$ is $\paramc_\Lambda/2$--$K$ non--resonant modulo $\Lambda$.
\end{lemma}

\proof of (i): 
Let us first consider $I\in \paramD_0$, so that $I\notin {\mathcal Z}_1$. For 
any $k\in {\mathbb Z}^n$, with $\norm{k}\leq K$, let us denote by $\tilde k$ 
the vector which generates the maximal one dimensional $K$--lattice
containing $k$. Since $I\notin {\mathcal Z}_1$ we have:
$$
\Norm{\pi_{\langle \tilde k \rangle}\omega(I)} \geq {\lambda_1\over \Norm{\tilde k}} \geq 
{\lambda_1\over \Norm{k}} ,
$$
and consequently $\norm{k\cdot \omega(I)} = \Norm{k}\Norm{\pi_{\langle \tilde k \rangle}\omega(I)} \geq \lambda_1$.  
Therefore, $\paramD_0$ is $\lambda_1$--$K$ non resonant.\\

\giu
Now,  consider a maximal $K$--lattice $\Lambda$, with 
$j:=\dim \Lambda \in \{1,\ldots ,n-1\}$ and let $I\in \paramD_\Lambda$.
As  in \cite{poschel},  let $k\notin \Lambda$  with $\norm{k}\leq K$ and  denote by $\Lambda_+$ the  maximal $K$--lattice generated by $\Lambda$ and $k$ (since $\Lambda$ is maximal, 
$\dim \Lambda_+=j + 1$). For the purpose of this proof, let us denote 
$$\pi:=\pi_{\langle\Lambda\rangle}\ ,\quad \pi_\perp:=\pi_{\langle \Lambda\rangle^\perp}=
\idmap-\pi\ ,\qquad  \pi_+:=\pi_{\langle\Lambda_+\rangle}\ ,
$$
where $\idmap$ denotes the identity map. 
Since $\pi \pi_+=\pi$, it is easy to check that
$$\pi_\perp k \cdot \big(\pi_+ \omega(I)- \pi \omega(I)\Big) + \pi k\cdot \pi \omega(I)=k\cdot \omega(I)\ .
$$
Thus, since  
the vectors  $\pi_\perp k$ and $\pi_+\omega(I)-\pi\omega(I)=\pi_\perp \pi_+\omega(I)$ are proportional, and $|\Lambda_+|\le |\Lambda|\,  \|\pi_\perp k\|$, 
we obtain
\begin{eqnarray*}
\norm{k\cdot \omega(I)} &\ge & 
\big| \pi_\perp k \cdot \big(\pi_+ \omega(I)- \pi \omega(I)\big) \big| - \big|\pi k\cdot \pi \omega(I)
\big|\\
& =&
\| \pi_\perp k\|\, \|\pi_+ \omega(I)- \pi \omega(I)\|  - \big|\pi k\cdot \pi \omega(I)
\big|
\\
&\ge&{\norm{\Lambda_+}\over \norm{\Lambda}} 
\sqrt{\Norm{\pi_+\omega(I)}^2-
\Norm{\pi \omega(I)}^2}-
\Norm{\pi k}
\Norm{\pi \omega(I)} .
\end{eqnarray*}
Using $\Norm{\pi k}\leq \Norm{k}\leq  \norm{k}\leq K$, 
$\Norm{\pi \omega(I)}<
\paramp_\Lambda$, $\Norm{\pi  \omega(I)}\geq \paramp_{\Lambda_+}$ we 
obtain:
$$
\norm{k\cdot \omega(I)} \geq {\norm{\Lambda_+}\over \norm{\Lambda}} 
\sqrt{{\lambda_{j+1}^2\over \norm{\Lambda_+}^2}-
{\lambda_{j}^2\over \norm{\Lambda}^2}}-K\paramp_\Lambda\ \ .
$$
Using again $\norm{\Lambda_+}\leq \norm{\Lambda}K$, and $K\leq K^{a_j}$,
we obtain:
\begin{eqnarray}
\norm{k\cdot \omega(I)} &\geq& {1\over \norm\Lambda}\Big (\sqrt{ \lambda_{j+1}^2-
K^2\lambda_j^2}-K \lambda_j\Big )\geq 
\paramp_\Lambda  \Big (\sqrt{\Big ({\lambda_{j+1}\over \lambda_j}\Big )^2
-K^2}-K\Big )\cr
&\geq& \paramp_\Lambda 
 \Big (\sqrt{ (AK)^{2 a_j}-K^2}-K \Big )
\geq \paramp_\Lambda
 \left (\sqrt{ (AK)^{2 a_j}-K^{2a_j}}-K^{a_j} \right )\cr
 &=&
\paramp_\Lambda K^{a_j}  \Big (\sqrt{ A^{2a_j} -1}-1 \Big ) ,
\end{eqnarray}
so that, by (\ref{conda1}), we finally get:
$$
\norm{k\cdot \omega(I)} \geq \paramp_\Lambda K^{a_j} 
 \Big (\sqrt{ A^{2a_j} -1}-1 \Big )\geq E^{a_j} K^{a_j} \paramp_\Lambda=\paramc_\Lambda .
$$
\hfill $\Box$
 
\proof of (ii): If $j=n-1$, the conclusion follows directly from 
lemma \ref{L:geoml}-(i) and 
\equ{paramDn}.
Let us therefore 
consider, for any $j=1,\ldots ,n-2$, $I\in \paramD_{\Lambda,\paramk_\Lambda}^\paramh$ and 
$I'\in \paramD_\Lambda\cap (\paramD-\paramh)$ such that 
$I\in {\cal C}_{\Lambda,\paramk_\Lambda}^\paramh(I')$. 
By (\ref{conditionsin.2}) and  \equ{diamcilindr}  we get
$$
\Norm{I'-I}\leq  \paramk_\Lambda  + 4 \Big ({2\parame +M\paramb  \over \paramom}\,
{\paramp_\Lambda+M\paramk_\Lambda  \over C_j}\Big )^{1\over \parama_j} .
$$
Using also lemma \ref{L:geoml}-(i), for any $k\in {\mathbb Z}^n\backslash \Lambda$ 
with  
$\norm{k}\leq K$, we have 
\beq{KKK}
\norm{k\cdot \omega(I)}\geq \norm{k\cdot \omega(I')}- KM \Norm{I-I'}\geq \paramc_\Lambda - 
KM \Big (\paramk_\Lambda  + 4 \Big ({2\parame +M\paramb  \over \paramom}
{\paramp_\Lambda+M\paramk_\Lambda  \over C_j}\Big )^{1\over \parama_j}\Big ) .
\eeq
But, since, by \equ{conditionsin.1} and \equ{alphalambda},  $KM \paramk_\Lambda \leq  \paramc_\Lambda/ 4$ 
from  (\ref{conditionsin.3}) there follows
\begin{eqnarray}
4KM \Big ({2\parame +M\paramb  \over \paramom} 
{\paramp_\Lambda+M\paramk_\Lambda \over C_j} \Big )^{1\over \parama_j} & \leq & 
4KM \Big (2{2\parame +M\paramb  \over \paramom} 
{\paramp_\Lambda\over C_j} \Big )^{1\over \parama_j}\cr
&\leq& 
KM \paraml_j  \Big ({\paramp_\Lambda\over C_j} \Big )^{1\over \parama_j} \leq
{\paramc_\Lambda \over 4}  ,
\end{eqnarray}
which, together with \equ{KKK} yields $\norm{k\cdot \omega(I)}\geq \paramc_\Lambda/2$. 
\hfill $\Box$

\giu

\noindent
$\bullet$ \underline{\sl  Non overlapping of extended blocks and zones}

\begin{lemma} \label{L: geomnonov} For any maximal $K$--lattices 
$\Lambda\ne \Lambda'$ of 
the same dimension $j=1,\ldots ,n-1$,  we have 
$$\overline {\paramD^\paramh_{{\Lambda},\paramk_\Lambda}}\cap {\mathcal Z}_{\Lambda'}=\emptyset.$$
\end{lemma}
{\bf Proof.} Let $\Lambda\ne \Lambda'$ be 
maximal $K$--lattices of the same dimension $j\le n-1$ and consider  $I\in {\overline {\paramD_{{\Lambda},\paramk_\Lambda}^\paramh}}$: we have to prove that  
$I \notin {\mathcal Z}_{{\Lambda}'}$, i.e.,
\begin{equation}
\Norm{\pi_{\langle {\Lambda}'\rangle }\omega (I)} \geq  \paramp_{{\Lambda}'}  .
\label{toprove}
\end{equation}
We divide the proof in two steps: the case $j\le n-2$ and the case $j=n-1$.

\giu
{\bf Step 1.} ($1\le j\le n-2$). The argument follows from the following claims (i)$\div$(vi).

\giu
(i)  {\sl For any $\eta >0$, 
there exists  $I'\in \paramD_{\Lambda}\cap (\paramD-\paramh)$ 
such that}
$\Norm{I-I'}\leq  \paraml_j 
\Big ({\paramp_\Lambda \over C_j} \Big )^{1\over \parama_j}+ \eta$.

\giu
{\sl Proof of} (i):
Since $I\in {\overline {\paramD_{{\Lambda},\paramk_\Lambda}^\paramh}}$,
there exists $I''\in \paramD_{\Lambda,\paramk_\Lambda}^\paramh$ such that 
$\| I''-I\|<\eta$; (by definition of 
$\paramD_{\Lambda,\paramk_\Lambda}^\paramh$) there exists
$I'\in  \paramD_{\Lambda}\cap (\paramD-\paramh)$ such that
$I''\in {\discs}_{{\Lambda},\paramk_\Lambda}^\paramh(I')$. Then, 
(i) immediately follows from \equ{II''}.

\giu
(ii) 
$\Norm{\pi_{\langle \Lambda'\rangle}\omega(I')}\geq E^{a_j}K^{a_j-1}\paramp_\Lambda$.

\giu
{\sl Proof of} (ii):
Since $\Lambda\ne \Lambda'$, there exists $k\in \Lambda'$ such that 
$k\notin \Lambda$ and  $\norm{k}\leq K$. Therefore we have $\Norm{\pi_{\langle\Lambda'\rangle}
\omega(I')}\geq {\norm{k\cdot \omega(I')}/ \Norm{k}}$
and since $I'\in \paramD_{\Lambda}$, (ii) follows from Lemma \ref{L:geoml}.

\giu
(iii) $\Norm{\pi_{\langle\Lambda'\rangle}\omega(I)}\geq {1\over 2} E^{a_j}K^{a_j-1}
\paramp_{\Lambda}$. 

\giu
{\sl Proof of} (iii):
Choose  $\eta \leq {\paramc_{\Lambda}\over 4KM}$.  Then, 
by using (\ref{conditionsin.3}), (i) and (ii), we obtain
\begin{eqnarray*}
\Norm{\pi_{\langle\Lambda'\rangle}\omega(I)} &\geq& 
\Norm{\pi_{\langle\Lambda'\rangle}\omega(I')} -M \Norm{I-I'}
\geq   E^{a_j}K^{a_j-1}\paramp_\Lambda - M\eta - M \paraml_j 
\Big ({\paramp_\Lambda \over C_j} \Big )^{1\over \parama_j}\cr
&\geq&  E^{a_j}K^{a_j-1}\paramp_\Lambda - {\paramc_\Lambda \over 2 K}
= {1\over 2}E^{a_j}K^{a_j-1}\paramp_\Lambda . 
\end{eqnarray*}
Now, observe  that, if we have ${1\over 2}E^{a_j}K^{a_j-1}\paramp_\Lambda \geq 
\paramp_{\Lambda'}$, then  (\ref{toprove}) follows at once. Therefore, {\sl let us henceforth assume that}
\begin{equation}  
{1\over 2}E^{a_j}K^{a_j-1}\paramp_\Lambda <  \paramp_{\Lambda'}  
\qquad {\rm i.e.}\qquad 
{\norm{\Lambda'}\over \norm{\Lambda}} < {2 \over  E^{a_j}K^{a_j-1}}\  .
\label{assumeqvol}
\end{equation}

\giu
(iv) 
$\Norm{\pi_{\langle\Lambda'\rangle}\omega(I')}\geq {A^{a_j}K^{a_j-1}}\paramp_{\Lambda'} - 
2\paramp_\Lambda$.

\giu
{\sl Proof of} (iv): Since $\Lambda\ne \Lambda'$, we consider  $k\in \Lambda$ such that 
$k\notin \Lambda'$ and $\norm{k}\leq K$, and we denote by $\Lambda''$ the 
maximal $K$--lattice of dimension $j+1$ which contains $\Lambda'$ and $k$. For the purpose of the proof of (iv) let us denote:
$$
\pi:=\pi_{\langle\Lambda\rangle}\ ,\qquad
\pi':=\pi_{\langle\Lambda'\rangle}\ ,\qquad
\pi'':=\pi_{\langle\Lambda''\rangle}\ .
$$
First,  since $I'\in \paramD_{\Lambda}$, we have
\begin{equation}
\Norm{\pi'\omega(I')}\geq 
\Norm{\pi'(\idmap-\pi)\omega(I')}
-\Norm{\pi'\pi\omega(I')}\geq 
\Norm{\pi'(\idmap-\pi)\omega(I')}-\paramp_\Lambda .
\label{calsigm}
\end{equation}
Then, since $I'\in \paramD_{\Lambda}$,  $I'\notin {\mathcal Z}_{\Lambda''}$ and 
we have 
\begin{equation}
\Norm{\pi''\omega(I')}\geq \paramp_{\Lambda''} .
\label{stimazona}
\end{equation}
Let us consider the vector $\nu = \pi_{\langle \Lambda'\rangle^\perp}k$. 
We remark that $\nu \in \langle \Lambda''\rangle \backslash 0 $. 
In fact, on the one hand $k \notin \langle\Lambda'\rangle$, so that 
$\nu\ne 0$; on the other hand, since 
$\nu =k- \pi_{\langle \Lambda'\rangle}k$ is the 
sum of $k\in \langle \Lambda\rangle$ and of $- \pi'k \in \langle \Lambda'\rangle$, we have also $\nu \in \langle \Lambda''\rangle$. 
Therefore, since $\nu $ is orthogonal 
to $\langle\Lambda'\rangle$, we have:
\begin{equation}
\pi_{\langle {\Lambda}''\rangle}  = \pi_{\langle \Lambda'\rangle}+
\pi_{\langle \nu\rangle} .
\label{sumr}
\end{equation}
Moreover, we have:
\begin{equation}
{\norm{\nu \cdot k} \over \Norm{\nu}}\geq {\norm{\Lambda''}\over 
\norm{\Lambda'}} .
\label{stimavol}
\end{equation}
In fact, on the one hand we have
$$
{\norm{\nu \cdot k} \over \Norm{\nu}}= 
{\norm{  \pi_{\langle\Lambda'\rangle^\perp}k \cdot k}\over \Norm{\pi_{\langle
\Lambda'\rangle^\perp}k}}= \Norm{\pi_{\langle\Lambda'\rangle^\perp}k}  ,
$$
on the other hand we have
$$
\Norm{\pi_{\langle\Lambda'\rangle^\perp}k} \geq {\norm{\Lambda''}\over 
\norm{\Lambda'}}  .
$$
>From (\ref{sumr}), we obtain:
\begin{eqnarray}
\Norm{\pi'(\idmap-\pi)\omega(I')}^2 &=&
\Norm{\pi'\pi''(\idmap-
\pi)\omega(I')}^2\cr
&=&\Norm{\pi''(\idmap-\pi)
\omega(I')}^2-
\Norm{\pi_{\langle\nu\rangle} \pi''(\idmap-\pi)\omega(I')}^2\cr
&=& \Norm{\pi''(\idmap-\pi)\omega(I')}^2-{\norm{\nu \cdot  \pi''(\idmap-
\pi)\omega(I')}^2\over 
\Norm{\nu}^2}  .
\label{proglamsp}
\end{eqnarray}
We notice that: 
\begin{equation}
\pi''(\idmap-\pi)\omega(I') \ne 0  .
\label{veqdiff0}
\end{equation}
In fact, first we have
\begin{eqnarray*}
\Norm{\pi''(\idmap-\pi)\omega(I')}
&\geq & \Norm{\pi''\omega(I')} - 
\Norm{\pi''\pi\omega(I')}
\geq \paramp_{\Lambda''}-\paramp_\Lambda\\
&\geq &
{\lambda_j \over \norm{\Lambda''}} 
\left ( A^{a_j}K^{a_j}- {\norm{\Lambda''} \over \norm{\Lambda}}\right )  ,
\end{eqnarray*}
then, using (\ref{stimavol}), (\ref{assumeqvol}), we obtain
$$
{\norm{\Lambda''} \over \norm{\Lambda}} =  
{\norm{\Lambda''} \over \norm{\Lambda'}}
{\norm{\Lambda'} \over \norm{\Lambda}}\leq \Norm{k} 
{\norm{\Lambda'} \over \norm{\Lambda}} < 
 {2 \Norm{k}   \over  E^{a_j}K^{a_j-1}} \leq  
{2 K   \over  E^{a_j}K^{a_j-1}}  ,
$$
and therefore we have:
$$
\Norm{\pi''(\idmap-\pi)\omega(I')}
> {\lambda_j \over \norm{\Lambda''}} 
\left ( A^{a_j}K^{a_j}- {2 K   \over  E^{a_j}K^{a_j-1}} \right ) .
$$
Finally, since $K\geq 1$, $a_j\geq 1$, and using also (\ref{conda1}), we have
\begin{eqnarray*}
\Norm{\pi''(\idmap-\pi)\omega(I')}
&>& {\lambda_j \over \norm{\Lambda''}}  
\left ( A^{a_j} K - {2 K   \over  E^{a_j}} \right )=
 {\lambda_j K \over \norm{\Lambda''}} \left ( A^{a_j}  - 
{2    \over  E^{a_j}} \right )
\\
&\geq& 
{\lambda_j K \over \norm{\Lambda''}} \left ( 2+ {2    \over  E^{a_j}} \right )>0  .
\end{eqnarray*}
Therefore, from (\ref{proglamsp}), (\ref{veqdiff0}), we have:
$$
\Norm{\pi'(\idmap-\pi)
\omega(I')} = 
\Norm{\pi''(\idmap-\pi)\omega(I')} 
\sqrt{1-{(\nu \cdot \pi'' (\idmap-\pi)\omega(I'))^2
\over \Norm{\nu}^2\Norm{\pi''(\idmap-\pi)\omega(I')}^2}} ,
$$
and, since 
$$
\pi''(\idmap-\pi)\omega(I') \cdot k=
(\idmap-\pi)\omega(I') \cdot k=0 ,
$$
we obtain:
\begin{equation}
\Norm{\pi'(\idmap-\pi)\omega(I')} \geq 
\Norm{\pi''(\idmap-\pi)\omega(I')} 
\min_{u \in k^{\perp}, \Norm{u}=1}
\sqrt{1-{(\nu \cdot u)^2\over \Norm{\nu}^2}} .
\label{minequal}
\end{equation}
We remark that the maximum of $\norm{\nu \cdot u}=\norm{\pi_{\langle 
k^{\perp}\rangle }\nu
\cdot u}$, for $u \in k^{\perp}$ and $\Norm{u}=1$,  
  is obtained for $u$ parallel to $\pi_{\langle k^\perp\rangle}\nu$, that 
is for   $u =\pi_{\langle k^{\perp}\rangle}\nu /\Norm{\pi_{\langle k^{\perp}\rangle }\nu}$. Therefore, we have: 
$$
\max_{u \in  k^{\perp}, \Norm{u}=1} 
\norm{\nu \cdot u}=\Norm{\pi_{\langle k^{\perp}\rangle }\nu}
$$
and correspondingly:
\begin{eqnarray*}
\min_{u \in  k^{\perp}, \Norm{u}=1} 
\sqrt{1-{(\nu \cdot u)^2\over \Norm{\nu}^2}}&=&
\sqrt{1- {\Norm{\pi_{\langle k^{\perp}\rangle}\nu}^2 \over \Norm{\nu}^2}}=
\sqrt{{\Norm{\nu}^2-\Norm{\pi_{\langle k^{\perp}\rangle}\nu}^2\over \Norm{\nu}^2}}=
{\Norm{\pi_{\langle k\rangle} \nu}\over \Norm{\nu}}\\
&=&
{\norm{\nu \cdot k}\over \Norm{\nu}\Norm{k}}  .
\end{eqnarray*}
Therefore, from (\ref{minequal}) and  (\ref{stimavol}) we obtain
$$
\Norm{\pi'(\idmap-\pi)\omega(I')} \geq
\Norm{\pi''(\idmap-\pi)\omega(I')} 
{\norm{\nu \cdot k}\over \Norm{\nu}\Norm{k}} 
\geq 
\Norm{\pi''(\idmap-\pi)\omega(I')} 
{\vert {\Lambda}''\vert \over \vert {\Lambda}'\vert\Norm{k}}  .
$$
Then, since: $\Norm{\pi''(\idmap-\pi)\omega(I')}\geq 
\Norm{\pi''\omega(I')}-\Norm{\pi''\pi
\omega(I')}\geq \paramp_{\Lambda''}-\paramp_\Lambda$, we obtain:
\begin{equation}
\Norm{\pi'(\idmap-\pi)\omega(I')} \geq  
(\paramp_{\Lambda''}-\paramp_\Lambda){\vert {\Lambda}''
\vert \over \vert {\Lambda}'\vert \Norm{k}}\geq 
{\lambda_{j+1}\over \vert {\Lambda}'\vert \Norm{k}}- 
\paramp_\Lambda {\vert {\Lambda}''
\vert \over \vert {\Lambda}'\vert \Norm{k}}\geq 
{A^{a_j}K^{a_j-1}}\paramp_{\Lambda'} - 
\paramp_\Lambda  ,
\end{equation}
and using  (\ref{calsigm}) we obtain (iv).

\giu
We now are ready to finish the proof of  (\ref{toprove}) in the case $j\le n-2$.
>From inequalities (iv) and (i), we obtain:
\begin{equation}
\Norm{\pi'\omega (I)}\geq 
 {A^{a_j}K^{a_j-1}}\paramp_{\Lambda'} - 
2\paramp_\Lambda-M \Big ( \paraml_j 
\Big ({\paramp_\Lambda \over C_j} \Big )^{1\over \parama_j}+ \eta\Big ) .
\label{stii}
\end{equation}
Using (\ref{conditionsin.3}) and choosing $\eta\leq {\paramc_{\Lambda}\over 4KM}$, we obtain:
\begin{equation}
\Norm{\pi'\omega (I)}\geq 
{A^{a_j}K^{a_j-1}}\paramp_{\Lambda'} - 2\paramp_\Lambda
-{1\over 2} E^{a_j}K^{a_j-1}\paramp_\Lambda .
\end{equation}
Since we are assuming \equ{assumeqvol}
and since $K^{a_j-1}\geq 1$, we obtain
\begin{eqnarray*}
\Norm{\pi'\omega (I)}&\geq &
A^{a_j}K^{a_j-1}\paramp_{\Lambda'} -2 \paramp_\Lambda 
-{1\over 2} E^{a_j}K^{a_j-1}\paramp_\Lambda\\
&>& A^{a_j}K^{a_j-1}\paramp_{\Lambda'}- {4\over E^{a_j} 
K^{a_j-1}} \paramp_{\Lambda'}-\paramp_{\Lambda'}  
>  \Big ( A^{a_j}-{4\over E^{a_j}}-1 \Big )\paramp_{\Lambda'} ,
\end{eqnarray*}
which, by (\ref{conda1}), yields (\ref{toprove}). 

\giu
{\bf Step 2.}
We now consider maximal $K$--lattices $\Lambda \ne \Lambda'$ of the 
same dimension $j=n-1$. Since $\paramD_{\Lambda,\paramk_\Lambda}^\paramh=\paramD_\Lambda \cap 
(\paramD-\paramh)$, we have $I \in {\overline \paramD_\Lambda}$ and
\begin{equation}
\Norm{\pi'\omega(I)}\geq E^{a_j}K^{a_j-1}\paramp_\Lambda  .
\label{picalmn}
\end{equation}
In fact, since $\Lambda\ne \Lambda'$, there exists $k\in \Lambda'$ such that 
$k\notin \Lambda$ and  $\norm{k}\leq K$. Therefore we have 
$\Norm{\pi' \omega(I)}\geq {\norm{k\cdot \omega(I)}/ \Norm{k}}$
and since $I\in {\overline \paramD_{\Lambda}}$, by lemma \ref{L:geoml} we have
$$
\Norm{\pi'\omega(I)}\geq {\norm{k\cdot \omega(I)}\over \Norm{k}}
\geq E^{a_j}K^{a_j-1}\paramp_\Lambda .
$$
We also have:
\begin{equation}
\Norm{\pi'\omega(I)}\geq {A^{a_j}K^{a_j-1}}\paramp_{\Lambda'} - 
2\paramp_\Lambda\ \ .
\label{stimmn}
\end{equation}
First,  since $I\in {\overline \paramD_{\Lambda}}$, we have
\begin{equation}
\Norm{\pi'\omega(I)}\geq 
\Norm{\pi'(\idmap-\pi)\omega(I)}
-\Norm{\pi'\pi \omega(I)}\geq 
\Norm{\pi'(\idmap-\pi)\omega(I)}
-\paramp_\Lambda .
\label{calsigman}
\end{equation}
Then, since $\Lambda\ne \Lambda'$, we consider  $k\in \Lambda$ such that 
$k\notin \Lambda'$ and $\norm{k}\leq K$. In 
particular, since $I\in {\overline \paramD_{\Lambda}}$, we have 
\begin{equation}
\Norm{\omega(I)}\geq \paramf .
\label{stimazonan}
\end{equation}
Let us consider the vector $\nu = \pi_{\langle \Lambda'\rangle^\perp}k$. 
Since $\nu $ is orthogonal to $\langle\Lambda'\rangle$, we have:
\begin{equation}
\idmap  = \pi'+\pi_{\langle \nu\rangle} .
\label{sumretn}
\end{equation}
Moreover, we have:
\begin{equation}
 \Norm{\pi_{\langle\Lambda'\rangle^\perp}k}={\norm{\nu \cdot k} \over \Norm{\nu}}\geq {1\over \norm{\Lambda'}} .
\label{stimavolumn}
\end{equation}
In fact, since the $K$--lattice  
$\langle \Lambda',k\rangle$ is generated by $\Lambda'$ and $k$
is properly contained in ${\mathbb Z}^n$, we have
$$
\norm{\Lambda'} \Norm{\pi_{\langle\Lambda'\rangle^\perp}k} \geq 
\norm{ \langle \Lambda',k\rangle} \geq 1  .
$$
>From (\ref{sumretn}), we obtain:
$$
\Norm{\pi(\idmap-\pi)\omega(I)}^2 =
\Norm{(\idmap-\pi)\omega(I)}^2-
\Norm{\pi_{\langle\nu\rangle} 
(\idmap-\pi)\omega(I)}^2 ,
$$
and since:
$$
\Norm{\pi_{\langle\nu\rangle} (\idmap-
\pi)\omega(I)}=
{\norm{\nu \cdot (\idmap-
\pi)\omega(I)}\over 
\Norm{\nu}} ,
$$
and:
$$
\Norm{(\idmap-\pi)\omega(I)}  \geq \paramf - 
\paramp_\Lambda \geq \paramf -\lambda_{n-1}>0  ,
$$
we have:
$$
\Norm{\pi'(\idmap-\pi)
\omega(I)} = 
\Norm{(\idmap-\pi)\omega(I)} 
\sqrt{1-{(\nu \cdot (\idmap-\pi)\omega(I))^2
\over \Norm{\nu}^2\Norm{(\idmap-\pi)\omega(I)}^2}} .
$$
Then, since $(\idmap-\pi)\omega(I) \cdot k=0$, we have
\begin{equation}
\Norm{\pi'(\idmap-\pi)\omega(I')} \geq 
\Norm{(\idmap-\pi)\omega(I')} 
\min_{u \in k^{\perp}, \Norm{u}=1}
\sqrt{1-{(\nu \cdot u)^2\over \Norm{\nu}^2}} .
\label{mineqn}
\end{equation}
We remark that the maximum of $\norm{\nu \cdot u}=\norm{\pi_{\langle 
k^{\perp}\rangle }\nu
\cdot u}$, for $u \in k^{\perp}$ and $\Norm{u}=1$,  
  is obtained for $u$ parallel to $\pi_{\langle k^\perp\rangle}\nu$, that 
is for   $u =\pi_{\langle k^{\perp}\rangle}\nu /\Norm{\pi_{\langle k^{\perp}\rangle }\nu}$. Therefore, we have: 
$$
\max_{u \in  k^{\perp}, \Norm{u}=1} 
\norm{\nu \cdot u}=\Norm{\pi_{\langle k^{\perp}\rangle }\nu}
$$
and correspondingly:
\begin{eqnarray*}
\min_{u \in  k^{\perp}, \Norm{u}=1} 
\sqrt{1-{(\nu \cdot u)^2\over \Norm{\nu}^2}}&=&
\sqrt{1- {\Norm{\pi_{\langle k^{\perp}\rangle}\nu}^2 \over \Norm{\nu}^2}}=
\sqrt{{\Norm{\nu}^2-\Norm{\pi_{\langle k^{\perp}\rangle}\nu}^2\over \Norm{\nu}^2}}=
{\Norm{\pi_{\langle k\rangle} \nu}\over \Norm{\nu}}\\
&=&
{\norm{\nu \cdot k}\over \Norm{\nu}\Norm{k}}  .
\end{eqnarray*}
Therefore, from (\ref{mineqn}), we obtain:
$$
\Norm{\pi'(\idmap-\pi)\omega(I)} \geq 
\Norm{(\idmap-\pi)\omega(I)} 
{\norm{\nu \cdot k}\over \Norm{\nu}\Norm{k}}  ,
$$
and from (\ref{stimavolumn}) we obtain also:
$$
\Norm{\pi'(\idmap-\pi)\omega(I)} \geq 
\Norm{(\idmap-\pi)\omega(I)} 
{ 1 \over \vert {\Lambda}'\vert\Norm{k}}  .
$$
Then, since: $\Norm{(\idmap-\pi)\omega(I)}\geq 
\Norm{\omega(I)}-\Norm{\pi
\omega(I)}\geq \paramf-\paramp_\Lambda$, we obtain:
\begin{equation}
\Norm{\pi'(\idmap-\pi)\omega(I)} \geq  
(\paramf-\paramp_\Lambda){1 \over \vert {\Lambda}'\vert \Norm{k}}\geq 
{A^{a_j}K^{a_j-1}}\paramp_{\Lambda'} - 
\paramp_\Lambda  ,
\end{equation}
and using  (\ref{calsigman}) we obtain (\ref{stimmn}).
\vskip 0.4 cm
\noindent
If $ E^{a_j}K^{a_j-1}\paramp_\Lambda \geq 2 \paramp_{\Lambda'}$, using (\ref{picalmn}),
there is nothing more to prove. Therefore, we assume:
$$
E^{a_j}K^{a_j-1}\paramp_\Lambda <  2 \paramp_{\Lambda'} \ .
$$
Then, using (\ref{stimmn}), we obtain:
\begin{equation}
\Norm{\pi'\omega (I)}\geq  
A^{a_j}K^{a_j-1}\paramp_{\Lambda'} - \paramp_\Lambda 
>A^{a_j}K^{a_j-1}\paramp_{\Lambda'}- {2\over E^{a_j} 
K^{a_j-1}} \paramp_{\Lambda'}  .
\end{equation}
Since $K^{a_j-1}\geq 1$, we have:
$$
\Norm{\pi'\omega (I)}>    
\Big ( A^{a_j}-{2\over E^{a_j}} \Big )\paramp_{\Lambda'} ,
$$
and using (\ref{conda1}) we obtain (\ref{toprove}).

\hfill $\Box$

\section{Normal forms and dynamics in resonant blocks}
The geometric construction of \S~\ref{sec:geometry} together with normal form theory allows to have
some control of the dynamics in the extended blocks. 
We shall use normal form theory in the version given by 
P\"oschel in \cite{poschel}; see, in particular, the  ``Normal Form Lemma" at p.~192 of \cite{poschel} (which  we shall use   with 
parameters $p=q=2$); notice that the constant $M$ used in \cite{poschel} is an upper bound on the 
derivative of $\omega(I)$, which is used only as Lipschitz constant, so that our notation is consistent with that used in \cite{poschel}. 

\noindent
In fact, the following lemma holds

\begin{lemma}
\label{dynamics}
{\rm (i)} Let  $(I_t,\varphi_t)$ be the solution of the Hamilton equations 
with  initial condition\footnote{I.e. $I_t|_{t=0}=\bar I_0$:
we are using here a slight abuse of notation in order not to confuse
the point $I_0$ in the statment of Theorem~1 with the arbitrary point $\bar I_0$ used here.} 
$(\bar I_0,\varphi_0)\in \paramD_0^\paramh\times {\mathbb T}^n$. Then, 
\begin{equation}
\Norm{I_t - \bar I_0} \leq \paramn 
\label{nonrest}
\end{equation}
for all times\footnote{Recall the definition of $T_0$ in \equ{TT}.} $|t|\le T_0$.
\\
{\rm (ii)}
Let   $\Lambda$ be a maximal $K$--lattice  of dimension $j\in \{1,\ldots,n-1\}$, and let  $(I_t,\varphi_t)$ be the solution of the Hamilton 
equations with initial data  $(\bar I_0,\varphi_0)\in (\paramD_{\Lambda}\cap (\paramD-(j+1)\paramh))\times
{\mathbb T}^n$. 
Let $\tau_{\!\rm e}$ be the (possibly  infinite) exit time  from\footnote{I.e., $\tau_{\!\rm e}$ is such that $I_t \in \paramD_{\Lambda,\paramk_\Lambda}^\paramh$ for 
$ |t| <|\tau_{\!\rm e}|$ and $I_{\tau_{\!\rm e}} \notin  \paramD_{\Lambda,\paramk_\Lambda}^\paramh$.} $ \paramD_{\Lambda,\paramk_\Lambda}^\paramh$.
Then, if   $|{\tau_{\!\rm e}}| \geq T_\Lambda$, 
we have $\Norm{I_t -\bar I_0} \leq r_j$ for any time $ |t| < T_\Lambda$;
otherwise, there exists $0\le i\le j-1$ such that $I_{\tau_{\!\rm e}}\in \paramD_i\cap(\paramD-j\paramh)$.
\end{lemma}

\proof of (i): 
The non--resonant block $\paramD_0$ is $\lambda_1$ non--resonant (see Lemma 
\ref{L:geoml}). Let us consider as extension vector $(\paramn,\paramg )$. 
Because of the definition of $\paramn$, \equ{rho0}, \equ{Ksigma} and the first inequality in \equ{epsnonres}, 
we can apply the normal form lemma 
in \cite{poschel}  in $\paramD_0$. It then follows at once \equ{nonrest}
for all times $|t|\le T_0$  with $T_0$ as in \equ{TT}. 
\hfill $\Box$

\vskip0.5cm
\proof of (ii): Let us first assume that $|{\tau_{\!\rm e}}|\ge T_\Lambda$ and consider 
the extension vector $(\paramb_{\Lambda},
\paramg )$.
By Lemma~\ref{L:geoml}--(ii), 
the domain $\paramD_{\Lambda,\paramk_\Lambda}^\paramh$ is 
$\paramc_\Lambda/2$--$K$ non--resonant modulo $\Lambda$. 
Thus, since $\paramb_{\Lambda}\leq \paramb$ (by \equ{rhoLrho}), and because of 
\equ{Ksigma}, the definition of $\paramd_\Lambda$ and 
the third inequality in \equ{epsnonres}, 
 we can apply the Normal Form Lemma  in \cite{poschel} (with $p=q=2$), 
in $\paramD_{\Lambda,\paramk_\Lambda}^\paramh$.
Thus, there exists   a canonical transformation:
\begin{eqnarray}
\phi: &(\paramD_{\Lambda,\paramk_\Lambda}^\paramh)_{{\paramb_{\Lambda}\over 2}}
\times {\mathbb T}^n_{{\paramg \over 6}} \rightarrow 
(\paramD_{\Lambda,\paramk_\Lambda}^\paramh)_{\paramb_{\Lambda}}
\times {\mathbb T}^n_{\paramg }&\cr
&(I',\varphi') \longmapsto (I,\varphi)=\phi(I',\varphi')
\end{eqnarray}
conjugating $H$ to its resonant normal form:
\begin{equation}
H_\Lambda = H\circ \phi= h+\parameps g+\parameps f_*
\label{hlam}
\end{equation}
with $g$ a real--analytic functions having the Fourier expansion
\beq{exp g}
g=\sum_{k\in \Lambda} g_k \exp(i k\cdot \varphi)\ ,
\eeq
and the ``remainder'' $f_*$ satisfying the exponential bound:
\begin{equation}
\norm{f_*}_{{ \paramD_{\Lambda,\paramk_\Lambda}^\paramh};{\paramb_{\Lambda}\over 2},{\paramg \over 6}}
\leq e^{-K {\paramg \over 6}}\norm{f}_{\paramb,\paramg}   .
\label{normf}
\end{equation}
Also, for any $(I',\varphi')\in 
(\paramD_{\Lambda,\paramk_\Lambda}^\paramh)_{{\paramb_{\Lambda}\over 2}}\times {\mathbb
  T}^n$, by the third inequality in \equ{epsnonres}, one has:
$$
\Norm{I'-I}\leq {8K \over \paramc_\Lambda}\parameps \norm{f}_{\paramb,\paramg} 
\leq {1\over 2^6}  \paramb_{\Lambda}   , 
$$
so that $\phi^{-1}(\paramD_{\Lambda,\paramk_\Lambda}^\paramh\times {\mathbb T}^n
)\subseteq (\paramD_{\Lambda,\paramk_\Lambda}^\paramh)_{{\paramb_{\Lambda}\over 2^6}}\times 
{\mathbb T}^n$. Finally,
using the second inequality in \equ{epsnonres}
we have also:
$$
\Norm{I'-I}\leq {8K \over \paramc_\Lambda}\parameps \norm{f}_{\paramb,\paramg} 
\leq {1\over 2^6}\paramk_\Lambda \   .
$$
Therefore, since $I_t\in \paramD_{\Lambda,\paramk_\Lambda}^\paramh$ for any $|t| <|{\tau_{\!\rm e}}|$, 
we may define $(I'_t,\varphi'_t)=\phi^{-1}(I_t,\varphi_t)$,
and using the specific form of hamiltonian (\ref{hlam}), we have
$$
\Norm{\pi_{\langle\Lambda^{\perp}\rangle} (I'_t-I'_0)} 
\le \parameps
\Norm{\int_0^t {\partial f_*\over \partial \varphi}(I'_t,\varphi'_t)dt}
\leq \parameps\norm{t} 
\sup_{(\paramD_{\Lambda,\paramk_\Lambda}^\paramh)_{{\paramb_{\Lambda}\over 2^6}}\times {\mathbb T}^n}
\Norm{{\partial f_*\over \partial \varphi}}  .
$$
By Cauchy estimate (see Lemma B.3 of \cite{poschel}) and by (\ref{normf}), 
we have:
$$
\sup_{(\paramD_{\Lambda,\paramk_\Lambda}^\paramh)_{{\paramb_{\Lambda}\over 2^6}}\times {\mathbb T}^n}
\Norm{{\partial f_*\over \partial \varphi}} \leq 
{6\over e s} 
\norm{f_*}_{{ \paramD_{\Lambda,\paramk_\Lambda}^\paramh};{\paramb_{\Lambda}\over 2},{\paramg \over 6}}
\leq  {6 \over e\paramg }e^{-K {\paramg \over 6}}
\norm{f}_{\paramb,\paramg}  ,
$$
so that, for any $|t|< T_\Lambda$, we have:
$$
\Norm{\pi_{\langle\Lambda^{\perp}\rangle} (I'_t-I'_0)}  \leq 
{6\parameps\over e\paramg }\norm{t}e^{-K {\paramg \over 6}}\norm{f}_{\paramb,\paramg}\leq {1\over 4}\paramk_\Lambda .
$$
As a consequence,  the 
motion $I_t$ has the representation:
\begin{equation}
I_t =\bar I_0+v(t)+d(t)
\label{fdrep}
\end{equation}
with $v(t)\in \langle\Lambda\rangle$ with $v(0)=0$, and 
$\Norm{d(t)}<{3\over 4}\paramk_\Lambda$: indeed, we can write
$$
I_t=\bar I_0+(I_t-I'_t)+(I'_t-I'_0)+(I'_0-\bar I_0) ,
$$
and take $v(t)=\pi_{\langle\Lambda\rangle} (I'_t-I'_0)$ and $d(t)=(I_t-I'_t)+
\pi_{\langle\Lambda\rangle^\perp} (I'_t-I'_0)+(I'_0-\bar I_0)$.

\noindent
Therefore,  $I_t \in \paramD_{\Lambda,\paramk_\Lambda}^\paramh 
\subseteq {\cal  Z}_\Lambda \cap (\paramD-\paramh) $ and  because 
of the representation (\ref{fdrep}) the distance between $I_t$ and the space 
$\bar I_0+\langle\Lambda\rangle$ is smaller than ${3\over 4}\paramk_\Lambda$.
Furthermore, $I_t$ is connected 
to $\bar I_0$ in the set 
$\Big (\cup_{I'\in \bar I_0+\langle\Lambda\rangle} \B(I',{3\over 4}\paramk_\Lambda)\Big )\cap 
{\mathcal Z}_\Lambda\cap (\paramD-\paramh)$ so that  $I_t \in 
{\cal C}_{\Lambda,{3\over 4}\paramk_\Lambda}^{\paramh}(\bar I_0)\subset 
{\cal C}_{\Lambda,\paramk_\Lambda}^{\paramh}(\bar I_0)$. Thus, by Lemma~\ref{lem:actions} we have $\|I_t-\bar I_0\|\le r_j$ for any $|t|< T_\Lambda$, as claimed.

\vskip 0.4 cm
\noindent
Let us now assume that the exit time ${\tau_{\!\rm e}}$ satisfies: $0< |{\tau_{\!\rm e}}| < T_\Lambda$. 
Since for any time $|t| <|\tau_{\!\rm e}|$, we have
$I_t \in {\cal C}^{\paramh}_{\Lambda,{3\over 4}\paramk_\Lambda}(\bar I_0)$, we have 
also: $I_{\tau_{\!\rm e}} \in {\overline {{\cal C}^{\paramh}_{\Lambda,{3\over
  4}\paramk_\Lambda}(\bar I_0)}}$. As a consequence (again Lemma~\ref{lem:actions}), we have $\Norm{I_t-\bar I_0} \leq r_j  
< \paramh$ for any $|t|\le |{\tau_{\!\rm e}}|$and since $\bar I_0\in \paramD-(j+1)\paramh$, we also have 
\beq{I delta}
I_{\tau_{\!\rm e}}\in \paramD-j\paramh\ .
\eeq
Since $I_t \in {\cal C}^{\paramh}_{\Lambda,{3\over
    4}\paramk_\Lambda}(\bar I_0)$, the distance between 
$I_t$ and $\bar I_0+\langle\Lambda\rangle$ is strictly smaller than ${3\over
  4}\paramk_\Lambda$, and the distance between  $I_{\tau_{\!\rm e}}$ and $\bar I_0+\langle\Lambda\rangle$
is smaller or equal than  ${3\over 4}\paramk_\Lambda$. Finally, 
since $I_t \in {\cal C}^{\paramh}_{\Lambda,{3\over
    4}\paramk_\Lambda}(\bar I_0)$, we have $I_t \in {\mathcal Z}_\Lambda$, that is:
$\Norm{\pi_{\langle\Lambda\rangle} \omega(I_t)}<\paramp_\Lambda$. As a consequence, 
since $I_{\tau_{\!\rm e}} \notin \paramD_{\Lambda,\paramk_\Lambda}^\paramh$, the only possibility is:
$\Norm{\pi_{\langle\Lambda\rangle} \omega(I_{\tau_{\!\rm e}})} = \paramp_\Lambda$.
But this means that $I_{\tau_{\!\rm e}}\notin {\mathcal Z}_\Lambda$. On the other hand, by Lemma~\ref{L: geomnonov},
$I_{\tau_{\!\rm e}}$ cannot belong to any ${\mathcal Z}_{\Lambda'}$ for any maximal $K$--lattice  $\Lambda'\neq \Lambda$  of the same dimension $j$; therefore $I_{\tau_{\!\rm e}}\notin {\mathcal Z}_j$, whence, by \equ{Dj}, there must exist an $i\in {0,...,j-1}$ such that $I_{\tau_{\!\rm e}}\in \paramD_i$, which, together with \equ{I delta}, concludes the proof of the lemma.
\hfill$\Box$

\section{The resonance trap argument and conclusion of the proof}
We are now in position to conclude the proof of the theorem, proving \equ{main 1} and \equ{main 2}.

\vskip0.5cm
\noindent
In view of \equ{covering},
the are are two alternatives\footnote{Recall that $I_0$ is the center of $\paramD\subset \paramB
$, which is a sphere  of radius $r=2\paramo \parameps^b=2n\paramh$.}:

\begin{itemize}
\item[(a)] either $I_0\in \paramD_0\cap (\paramD-n\paramh)$;

\item[(b)] or $I_0\in \paramD_\Lambda\cap (\paramD-(j+1)\paramh)$ for some maximal $K$--lattice of dimension $j\in \{1,...,n-1\}$.

\end{itemize}
In case (a), by Lemma~\ref{dynamics}--(i), by \equ{rho est},
the definition of $T_0$ and $T_{\rm exp}$ (\equ{TT}, \equ{TTT}) and \equ{Texp}, the theorem is proved.

\noindent
In case (b), by Lemma~\ref{dynamics}--(ii), there are two alternatives:

\begin{itemize}
\item[(b1)] either $\|I_t-I_0\|\le r_j\lepereq{rho est} \paramh$, for $|t|\le T_{\exp}\le T_{\Lambda}$

\item[(b2)] or there exist a $t_1$ such that $\|I_t-I_0\|\le \paramh$ for all $|t|\le |t_1|$ and $I_{t_1}\in \paramD_i\cap (\paramD-j\paramh)$ for some $i\in \{0,...,j-1\}$.
\end{itemize}
In case (b1), by \equ{TT}, \equ{TTT} and \equ{Texp}, and recalling that $\paramh= \paramo \parameps^b/n$, the theorem is proved.

\noindent
In case (b2) we iterate the above scheme. Hence, after $0\le k\le n-1$ steps we see that the action--trajectory
$I_t$ ends up  either  in a ``trapping resonant region'' $\paramD_\Lambda\subset \paramD_i$ where it gets stuck for exponentially long times
or it will end up in $\paramD_0$ where also gets stucked for exponentially long time. Since in such $k$ steps $I_t$ moves at most by $k\paramh$ we see that in the (possible) fast drift we have $\|I_t-I_0\|\le k\paramh \le (n-1)\paramh$ to which we have to add the displacement in the trapping region which is again at most $\paramh$. Thus, for times $|t|\le T_{\rm exp}$ we have $\|I_t-I_0\|\le n\paramh = \paramo\parameps^b$ as claimed. 
\hfill$\Box$

\vskip 0.5cm
\noindent
We remark that, in the case (b) above, $I_t$ may visit several blocks in 
the time $T_{\rm exp}$; let us denote by $j^*$ their minimal multiplicity, and $t^*<T_{\rm exp}$ be such that 
$I_{t^*} \in B_{\Lambda^*}\cap (B-(j^*+1)\rho)$ with $\dim \Lambda^*=j^*$. Then, we have 
$I_0 \in B_{\Lambda^*}$ (and, therefore, $I_t\in B_{\Lambda^*}$ for all $|t|\le T_{\rm exp}$). In fact, since the geometry of resonances of the 
Hamilton function $-H$ is identical to the  geometry of resonances of $H$, if 
we consider the solution $(I'_t,\varphi'_t)$ of the Hamilton 
equations of $-H$ with $I'_0=I_{t^*}$, and apply Lemma (\ref{dynamics}), we obtain that $I_0=I'_{-t*}\in B_\Lambda 
\cap B_{\Lambda^*}$.

\appendix
\small
\section{Angles between linear spaces}\label{appendix angles}
\setcounter{equation}{0}
\renewcommand{\theequation}{\ref{appendix angles}.\arabic{equation}}

\noi
In this appendix $n\ge 2$, $u,v, w, z...$ denote vectors in $\rn$ and $L_1, L_2, L',...$ linear vector subspaces of $\rn$ of dimension $m\in\{1,...,n-1\}$; $\pL$ denotese the orthogonal projection onto the linear space $L$ and
$\Arccos: [-1,1]\to [0,\pi]$ denotes the principal branch of the inverse real cosine.

\dfn{def1} Let .
The angle between $u$ and $v$  is defined as
\beqno
u\angolo v\:=\casesalt{\dst \Arccos \frac{u\cdot v}{\|u\|\, \|v\|}\ , \ \ \ \  }{u,v\neq 0}{\frac\p2\ .}
\eeqno
\edfn

\dfn{def2}
The angle between $L_1$ and $L_2$ is defined as
\beqno
L_1\angolo L_2:= \max_{u\in L_1\bks\{0\} }  u \angolo \pLd  u\ .
\eeqno
\edfn
We, next, list a few elementary properties of angles between linear spaces, whose simple proof is left to the reader (for the proof of items x and xi , see, also, [Nekhoroshev79, p. 45]).

\begin{itemize}

\item[i.] $u\angolo v\in [0,\p]$ and $u\angolo v\in [0,\p/2]$ if and only if $u\cdot v\ge 0$;
$L_1\angolo L_2\in [0,\p/2]$.

\item[ii.]  $L_1\angolo L_2=\frac\p2$ 
if and only if\footnote{$L^\perp:=\{u\in\rn: u\cdot v=0\ ,\ \forall\, v\in L\}$.} 
$L_1\cap  L_2^\perp\neq \{0\}$.\\
$L_1\angolo L_2<\frac\p2$ 
if and only if 
$\{u\in L_1: \pLd u=0\}=\{0\}$.

\item[iii.]   $\dst u\angolo \pL u= \Arccos \frac{\|\pL u\|}{\|u\|}\ , \qquad \forall \ u\neq 0$.

\item[iv.]
$u\angolo  \pL u + u\angolo \pLp u= \frac\p2\  \ ,\qquad \forall\ u\neq 0$.

\item[v.] $ u\angolo \pL u = \min_{v\in L\bks \{0\}} u\angolo v$.

\item[vi.]
$\dst \cos L_1\angolo L_2= \min_{u\in L_1\bks\{0\}} \max_{v\in L_2\bks\{0\}} \frac{u\cdot v}{\|u\|\, \|v\|}$.

\item[vii.]  For any $u$ and $v$ one has\footnote{But,  in general, $L_1\angolo L_2\neq L_2\angolo L_1$: for example,  if $n=3$, $L_1=\{(0,t,t):t\in\real\}$
and $L_2=\{x_3=0\}$, then $L_1\angolo L_2= \p/4$, while $L_2\angolo L_1=\p/2$. }  $u\angolo v= v\angolo u$.

\item[viii.] If $u\neq 0\neq v$, $u\angolo v$ coincides with the (Euclidean) length of the shortest geodesic (equivalently, shortest curve) on the unit sphere
$S^{n-1}:=\{\x\in\rn: \|\x\|=1\}$ having as end--points the projections of $u$ and $v$ on $S^{n-1}$. 

\item[ix.] $\dst u\angolo v\le  u\angolo w + w\angolo v$.  Also: 
$\dst L_1\angolo L_2 \le L_1\angolo L_3 + L_3\angolo L_2$.

\item[x.] $L_1\angolo L_2= L_2^\perp \angolo L_1^\perp$.

\item[xi.] If $\dim L_1=\dim L_2$, then $L_1\angolo L_2= L_2\angolo L_1$.

\end{itemize}

\section{Parameter Relations
}\label{inequalities}
\setcounter{equation}{0}
\renewcommand{\theequation}{\ref{inequalities}.\arabic{equation}}
For completeness, in this appendix, we prove the elementary inequalities \equ{conda1}$\div$\equ{Texp}. 
Recall the definitions of the parameters given in \equ{qj}$\div$\equ{TTT}. \\
First, we observe that from these definitions and the hypothesis $0\le \parameps \le \paramepsu$, it follows easily:
\beqa{aux.B.4}
&& E\ge 4\ ; \quad A:=6E\ge 24\ ,\quad  
K:=\Big(\frac{\paramepso}{\parameps}\Big)^a\ge\Big(\frac{\paramepso}{\paramepsu}\Big)^a \ge 1\ ; \quad  
1\le |\Lambda |\le K^j \ ;
\\
&& \paramp_\Lambda\le \lambda_j\ ;\quad
\paramh= \frac{\paramb \mu_0}{K^{1/\parama_{n-1}}}\ ;
\quad r=2n\paramh\ ;
\\
\label{aux.B.2}
&&
\paramq_1+1= \frac1{2a}\ge n\ ;\quad
\paramq_j\ge 2 \quad (j\le n-2)
\ ;\quad
\paramq_j\Big(1-\frac1{\parama_j}
\Big)=a_j-1-j \Big(1-\frac1{\parama_j}
\Big).
\label{aux.B.1}
\eeqa
{\bf \equ{conda1}:} It follows immediately from \equ{aux.B.4}.

\giu
{\bf \equ{Ksigma}:} It follows from $\dst \frac{\paramepsu}{\paramepso} \le \Big(\frac{\paramg}{6}\Big)^{\frac1{a}}$.

\giu
{\bf \equ{conditionsin.1}:} To get the 1st inequality observe that $\paramq_j\ge 1\ge 1/\parama_{n-1}$, 
$\paramb \mu_0\ge \paramom/(24\sqrt2\, M)$ so that
\beqano
\paramk_\Lambda&=&\frac{\paramom}{2\sqrt2}\, \frac1{(AK)^{\paramq_j}}\, \frac1{|\Lambda|} \, \frac1{M}\le 
\frac{\paramom}{2\sqrt2}\, \frac1{A} \, \frac1{K^{\frac1{\parama_{n-1}}}}  \, \frac1{M}\lepereq{aux.B.4} 
\frac{\paramom}{48\sqrt2}\,  \frac1{K^{\frac1{\parama_{n-1}}}} \, \frac1{M}=
\frac12 \frac{\paramom}{24\sqrt2 M}\, \Big(\frac{\parameps}{\paramepso}\Big)^b
\\
&\le& \frac{\paramb \mu_0}{2}\, 
\Big(\frac{\parameps}{\paramepso}\Big)^b=\frac{\paramh}2\ .
\eeqano
As for the 2nd inequality we have: $\dst \paramk_\Lambda\le \paramk_\Lambda\, \frac{(EK)^{a_j}}{4K}=\paramd_\Lambda$.

\giu
{\bf \equ{conditionsin.2}}, {\sl first inequality}: Using: $\paramq_j\ge 2\ge 1/\parama_{n-1}$, \equ{aux.B.4} and $\mu_0\ge 1/(6^2 2 \sqrt2)>2/(\sqrt2 (24)^2)$, one finds
$$ \paramp_\Lambda= \frac{\paramom}{2\sqrt2} \, \frac1{(AK)^{\paramq_j}}\, 
\frac1{|\Lambda|}\le 
\frac{\paramom}{2\sqrt2} \, \frac1{24^2}\, \frac1{K^{\frac1{\parama_{n-1}}}}
\le \frac{\paramom}{4}\, \frac{\mu_0}{K^{\frac1{\parama_{n-1}}}} = \frac{\paramom\paramh}{\paramb\, 4}\ .
$$

\giu
{\bf \equ{conditionsin.2}}, {\sl second inequality}:
$\dst \paramp_\Lambda\lepereq{conditionsin.2} \frac{\paramom}{\paramb}\, \frac{\paramh}4\lepereq{conditionsin.1} \frac{\paramom}{2\paramb}\, (\paramh-\paramk_\Lambda)$.

\giu
{\bf \equ{conditionsin.2}}, {\sl third inequality}:  $\dst \paramp_\Lambda = \frac{\paramom}{2\sqrt2}\,
\frac1{(AK)^{\paramq_j}}\le\paramf$.

\giu
{\bf \equ{conditionsin.3}:} Using: the definitions given, the inequality $|\Lambda|\le K^j$ and last equality in \equ{aux.B.1}, one has:
\beqano
\frac{KM \paraml_j\Big(\frac{\paramp_\Lambda}{C_j}\Big)^{\frac1{\parama_j}}}{\frac14 \paramc_\Lambda}
&=& \Big(\frac1{C_j}\Big)^{\frac1{\parama_j}}
\
(4M\paraml_j)\ K^{1-a_j+\paramq_j\big(1-\frac1{\parama_j}\big)}\ 
|\Lambda|^{1- \frac1{\parama_j}} \ \Big(
\frac{2\sqrt2}{\paramom} A^{\paramq_j}\Big)^{1-\frac1{\parama_j}}\  \frac1{E^{a_j}}
\\
&\le& \big(\frac1{C_j}\big)^{\frac1{\parama_j}}
\
(4M\paraml_j)\ K^{\paramq_j\big(1-\frac1{\parama_j}
\big)+1 - a_j+\big(1-\frac1{\parama_j}
\big)}\ \ \Big(
\frac{2\sqrt2}{\paramom} A^{\paramq_j}\Big)^{1-\frac1{\parama_j}}\  \frac1{E^{a_j}}
\\
&=& \left( \frac{(4M\paraml_j)^{\parama_j}\  
6^{np_j-j)(\parama_j-1)}
}{C_j \ \Big(\frac{\paramom}{2\sqrt2}\Big)^{\parama_j-1}}\ \frac1{E^{\parama_j+j(\parama_j-1)}}\right)^{\frac1{\parama_j}} \le 1\ ,
\eeqano
where last inequality comes from the definition of $E$.

\giu
{\bf \equ{rhoLrho}:} Using: $\paramq_j \geq 1$, $\paramq_j-a_j \geq 0$,  $E\geq 1$, $a/b=\parama_{n-1}\geq 1$, one has:
\beqano
{\paramd_\Lambda \over \paramb} &=& {1\over 8\sqrt{2}} {1\over 6^{\paramq_j}} 
{\paramom \over M\paramb} {1\over K^{\paramq_j-a_j+1}}{1\over E^{\paramq_j-a_j} }
{1\over \norm{\Lambda}}
\leq {1\over 48\sqrt{2}}{\paramom \over M\paramb}{1\over K}
\lepereq{aux.B.4} 
 {1\over 48\sqrt{2}}{\paramom \over M\paramb}
\left ( {\paramepsu\over \paramepso}\right )^a
\\
&\lepereq{quantitative}&  {1\over 48\sqrt{2}}{\paramom \over M\paramb}
\Big( \min \big( 1 , \frac{6\sqrt2}{n}
{M\paramb\over \paramom}\big)\Big)^{a\over b}
\leq {1\over 8n}<1 .
\eeqano

\giu
{\bf \equ{epsnonres}}, {\sl first inequality:} Using: the definitions given, $\paramq_1+1=\frac1{2a}=np_1$, $\parameps K^{1/a}=\paramepso$, the definition of $\paramepso$ and $E$, one has:
\beqno
\frac{2^8 K}{\lambda_1\paramn}\, \parameps\, \norm{f}_{\paramb,\paramg}=
\frac{2^4}{6^{2np_1-3}}\, \frac1{E}\le \frac{2^4}{6^3}\, \frac14=\frac1{54}<1\ ,
\eeqno
where in the  inequality we used $n\ge 3$, $p_1\ge 1$ and $E\ge 4$.

\giu
{\bf \equ{epsnonres}}, {\sl second inequality:} By the first inequality in \equ{epsnonres}, we see that the second inequality is implied by 
$$
\frac1{2^8}{\lambda_1 \paramn\over K}
\leq \frac{\paramc_\Lambda \paramk_\Lambda}{2^9 K} \ .
$$
Now, using: the definitions given, $|\Lambda|\le K^j$, $\paramq_1-\paramq_j=n(p_1-p_j)+(j-1)\ge 0$ 
and the relation $a_j+1+2(\paramq_1-\paramq_j-j)=a_j-1+2n(p_1-p_j)\ge 0$, one has:
$$
\frac1{2^8}{\lambda_1 \paramn\over K}\, 
\frac{2^9 K}{\paramc_\Lambda \paramk_\Lambda} =
\frac1{E^{a_j}}\, \frac1{A^{2(\paramq_1-\paramq_j)}}
\frac{|\Lambda|^2}{K^{a_j+1+2(\paramq_1-\paramq_j)}}\le 
\frac1{E^{a_j}}\, \frac1{A^{2(\paramq_1-\paramq_j)}}
\frac1{K^{a_j+1+2(\paramq_1-\paramq_j-j)}}\le \frac1{E}<1.
$$

\giu
{\bf \equ{epsnonres}}, {\sl third inequality:} It follows immediately from the 2nd inequality in \equ{conditionsin.1} and from the second inequality in 
\equ{epsnonres}.

\giu
{\bf \equ{rho est}:}  
$\dst \frac{\paramn}{\paramh}= \Big(4\sqrt2\, \frac{M\paramb}{\paramom}\, \mu_0\, A^{\paramq_1} \, K^{\paramq_1-\frac1{\parama_{n-1}}}\Big)^{-1}<1$ since $\paramq_1\ge 2>1/\parama_{n-1}$, $K\ge 1$, and $\mu_0\ge  \frac{M\paramb}{\paramom} \frac{6\sqrt2}{n}$.

\giu

$\dst \frac{r_{n-1}}{\paramh}=K^{-\frac{\paramq_{n-1} -1}{\parama_{n-1}}}\, 
\frac{\paraml_{n-1}}{\paramb \mu_0}\, 
\Big( \frac{\paramom}{C_{n-1}}\, \frac1{2\sqrt2 A^{\paramq_{n-1}}}\Big)^\frac1{\parama_{n-1}}\le 
\Big(\frac{6E}{A}\Big)^\frac1{\parama_{n-1}}=1\ ,
$
where in the inequality we used $\paramq_{n-1}\ge 2>1/\parama_{n-1}$, $K\ge 1$,
$\dst \frac1{\mu_0}\le \frac{\paramb}{\paraml_{n-1}} \Big(\frac{12 \sqrt2 E C_{n-1}}{\paramom}\Big)^\frac1{\parama_{n-1}}$ and $\paramq_{n-1}\ge 1$.

\giu
Now, let $1\le j\le n-2$ and define $\parama_j':=\parama_j-1$, $b_j:= \frac{\paramq_j}{\beta_j}-1$ (recall \equ{par.aux}) and observe that
\beq{bj} 
b_j\ge \frac{\parama_j}{\beta_j}>0\ ,\qquad \frac1{\parama_j}-1+ \frac{\paramq_j \paramh'_j}{\parama_j\beta_j}=\Big(1-\frac1{\parama_j}\big) \, b_j\ .
\eeq
Observe also that from the definitions of $\paraml_j$, $E$ and $\mu_0$ it follows that
\beq{QEm} 
\paraml_j\ge \frac{\paramom}{M} \, \Big(\frac{C_j}{\paramom}\Big)^\frac1{\parama_j}=\frac{\paramom^{1- \frac1{\parama_j}} \, C_j^\frac1{\parama_j}}{M}
\ ,
\qquad
E\ge 
\left({(M\paraml_j)^{\parama_j} \over C_j  
{\paramom^{\parama_j'}}} \right)^{\frac1{\beta_j}}\ ,
\qquad 
\mu_0\ge \frac{\paramom}{\paramb \, 24\sqrt2 \, M}\ .
\eeq
Then, noticing also that $\dst \frac{24\sqrt2}{6^\frac{\paramq_j}{\parama_j}}\le \dst \frac{24\sqrt2}{6^2}<1$, we obtain
\beqano
\frac{r_j}{\paramh} &=& 
K^{-\frac{\paramq_j -1}{\parama_{n-1}}}\, 
\frac{\paraml_{j}}{\paramb \mu_0}\, 
\Big( \frac{\paramom}{C_j}\, \frac1{2\sqrt2 (6E)^{\paramq_j}}\Big)^\frac1{\parama_j}\le 
\frac{M}{\paramom}\, \paraml_j\,  \Big(\frac{\paramom}{C_j}\Big)^\frac1{\parama_j}\, 
\left( \frac{C_j \paramom^{\parama_j'}}{(M \paraml_j)^{\parama_j}}\right)^{\frac{\paramq_j}{\parama_j\beta_j}}
\\
&\ugpereq{bj}&\left(\frac{\paramom^{1-\frac1{\parama_j}}}{M}\, \frac{C_j^\frac1{\parama_j}}{\paraml_j}
\right)^{b_j}
\lepereq{QEm} 1\ .
\eeqano

\giu
{\bf \equ{rho0}:} Using $\paramq_1+1= \frac1{2a}$, $6E\ge 1$ and $\parameps\le \paramepsu\le \paramepso \cdot \Big(\min\big( 1 , \frac{6\sqrt{2}}{n}
{M\paramb\over \paramom}\big)\Big)^\frac1b$ one finds:

\nl

$\dst 
{\paramn\over \paramb}= 
{\paramom \over 4 \sqrt{2} (6E)^{np_1-1} M\paramb} 
\left ( {\parameps\over \paramepso}\right )^{1\over 2}
\leq {\paramom \over 4 \sqrt{2}   M\paramb} 
\min \left ( 1 , {6\sqrt{2}\over n}
{M\paramb\over \paramom}\right ) \le 1.$

\giu
{\bf \equ{rho0r}:} Since $\paramh=\paramr/(2n)$, \equ{rho0r} follows at one from \equ{rho est}.

\giu
{\bf \equ{rleqrho}:} Set $\dst x={n\over 6\sqrt{2}} {\paramom \over M\paramb}$, $\dst y={n \over 18\sqrt{2}} $,
$\dst z={4n \paraml_{n-1}\over \paramb} \Big ({\paramom \over 
12 \sqrt{2} E  C_{n-1}}\Big )^{1\over \parama_{n-1}}$. Then, 
from the definitions given and the hypothesis $\parameps\le \paramepsu$ it follows:
\beqano
\frac{2\paramr}{\paramb}= 4n\mu_0 \Big( \frac{\parameps}{\paramepso}\Big)^b \le 4n\mu_0 \Big( \frac{\paramepsu}{\paramepso}\Big)^b
=\max ( x,y,z)\min (x^{-1},y^{-1},z^{-1},1)\le 1.
\eeqano

\giu
{\bf \equ{Texp}:}
>From the definitions given (and since $\frac{e}{24}>\frac1{10}$) it follows:
\beqano
\dst T_{\rm exp} \frac{\sqrt\parameps}{T e^\frac{K\paramg}{6}}
&=&6 A^\frac1a\, \sqrt\frac{\paramepso}\parameps\, \min\left(
\frac1{10}\, \frac1{(AK)^{\paramq_1}}\, \frac1K\ ,\ \frac{e}{24} \min_{1\le j\le n-1\atop \Lambda: \dim \Lambda=j}
\Big(\frac1{|\Lambda|}\, \frac1{(AK)^{\paramq_j}}\Big)
\right)\\
&\ge& 
\frac35 \, A^\frac1a \, \sqrt\frac{\paramepso}\parameps\, \min_{1\le j\le n-1} \frac1{A^{\paramq_j}}\, \frac1{K^{\paramq_j+1}}
=\frac35 K A^{np_1+1}\ge \frac35 A^4>1\ .
\eeqano

\end{document}